\documentclass[11pt]{article}

\usepackage{latexsym,color,graphicx,amsmath,amssymb}

\def\reals{{\mathbb R}}
\def\eps{{\varepsilon}}
\def\bd{{\partial}}
\def\A{{\cal A}}

\def\F{{\cal F}}
\def\G{{\cal G}}

\def\R{{\cal R}}

\def\T{{\cal T}}
\def\H{{\cal H}}
\def\N{{\cal N}}
\def\W{{\cal W}}

\def\oX{{\overline{X}}}
\def\oT{{\overline{T}}}
\def\oZ{{\overline{Z}}}

\DeclareMathOperator{\polylog}{polylog}

\def\marrow{\marginpar[\hfill$\longrightarrow$]{$\longleftarrow$}}
\def\micha#1{\textsc{(Micha says: }\marrow\textsf{#1})}

\newtheorem{theorem}{Theorem}[section]
\newtheorem{lemma}[theorem]{Lemma}
\newtheorem{proposition}[theorem]{Proposition}
\newtheorem{corollary}[theorem]{Corollary}
 
\begin{document}

\title{On Ray Shooting for Triangles in 3-Space and Related Problems\thanks{%
Work by Esther Ezra has been partially supported by NSF CAREER under grant CCF:AF-1553354 
and by Grant 824/17 from the Israel Science Foundation.
Work by Micha Sharir has been partially supported by ISF Grant 260/18, by grant 1367/2016
from the German-Israeli Science Foundation (GIF), and by
Blavatnik Research Fund in Computer Science at Tel Aviv University.
A preliminary version of this work appears in the proceedings of the 37th Symposium on Computational Geometry.}}

\author{Esther Ezra\thanks{
School of Computer Science, Bar Ilan University, Ramat Gan, Israel;
{\sf ezraest@cs.biu.ac.il, https://orcid.org/0000-0001-8133-1335 }}
\and Micha Sharir\thanks{
School of Computer Science, Tel Aviv University, Tel Aviv Israel;
{\sf michas@tauex.tau.ac.il, http://orcid.org/0000-0002-2541-3763 }}
}



\maketitle

\begin{abstract}
We consider several problems that involve lines in three dimensions,
and present improved algorithms for solving them. The problems include
(i) ray shooting amid triangles in $\reals^3$, (ii) reporting
intersections between query lines (segments, or rays) and input triangles,
as well as approximately counting the number of such intersections,
(iii) computing the intersection of two nonconvex polyhedra, (iv) detecting, 
counting, or reporting intersections in a set of lines in $\reals^3$, and
(v) output-sensitive construction of an arrangement of triangles in three dimensions.

Our approach is based on the polynomial partitioning technique.

For example, our ray-shooting algorithm processes a set of $n$ 
triangles in $\reals^3$ into a data structure for answering ray 
shooting queries amid the given triangles, which uses $O(n^{3/2+\eps})$ 
storage and preprocessing, and answers a query in $O(n^{1/2+\eps})$ time,
for any $\eps>0$. This is a significant improvement over known results, 
obtained more than 25 years ago, in which, with this amount of storage, 
the query time bound is roughly $n^{5/8}$. The algorithms for the other 
problems have similar performance bounds, with similar improvements 
over previous results. 

We also derive a nontrivial improved tradeoff between storage and query time.
Using it, we obtain algorithms that answer $m$ queries on $n$ objects in 
\[
\max \left\{ O(m^{2/3}n^{5/6+\eps} + n^{1+\eps}),\; O(m^{5/6+\eps}n^{2/3} + m^{1+\eps}) \right\}
\]
time, for any $\eps>0$, again an improvement over the earlier bounds.
\end{abstract}


\section{Introduction}

In this paper we consider several algorithmic problems that involve, 
explicitly or implicitly, a finite set of lines in three dimensions. 
The main problems that we consider are:

\begin{description}
\item[(i)]
\emph{Ray shooting amid triangles in three dimensions.}
We have a set $\T$ of $n$ triangles in $\reals^3$, and our goal is to
preprocess $\T$ into a data structure that supports efficient ray-shooting
queries, each of which specifies a ray $\rho$ and asks for the first
triangle of $\T$ that is hit by $\rho$, if such a triangle exists.
\item[(ii)]
\emph{Intersection reporting, emptiness, and approximate counting queries amid triangles in three dimensions.}
For a set $\T$ of $n$ triangles in $\reals^3$, we want to
preprocess $\T$ into a data structure that supports efficient intersection
reporting (resp., emptiness) queries, each of which specifies a line, ray, or segment $\rho$ 
and asks for reporting the triangles of $\T$ that $\rho$ intersects 
(resp., determining whether such a triangle exists). We want 
the queries to be output-sensitive, so that their cost is a small 
(sublinear) overhead plus a term that is nearly linear in the output size $k$.
In the related problem of approximate counting queries, we want to
preprocess $\T$ into a data structure, such that given a query $\rho$ as above, it
efficiently computes the number of triangles of $\T$ that $\rho$ intersects, up
to some prescribed small relative error.
\item[(iii)]
Compute the intersection of two nonconvex polyhedra.
The complexity of the intersection can be quadratic in the 
complexities of the input polyhedra, and we therefore seek an output-sensitive
solution, where the running time is a small (subquadratic) overhead plus a term 
that is nearly linear in $k$, where $k$ is the complexity of the intersection.
\item[(iv)]
Detect, count, or report intersections in a set of lines in 3-space.
Again, in the reporting version we seek an output-sensitive solution,
as above.
\item[(v)]
Output-sensitive construction of an arrangement of triangles in three dimensions.
\end{description}

All these problems, or variants thereof, have been considered in 
several works during the 1990s; see~\cite{AM,AgS,dB,BHOSK,CEGSS,MS,Pel} 
for a sample of these works.  See also Pellegrini~\cite{Pel:surv} for 
a recent comprehensive survey of the state of the art in this area.

Pellegrini~\cite{Pel} presents solutions to some of these problems,
including efficient data structures (albeit less efficient than ours)
for the ray-shooting problem, and also (a) an output-sensitive algorithm 
for computing the intersection of two nonconvex polyhedra in time 
$O(n^{8/5+\eps} + k\log k)$, for any $\eps>0$, where $n$ is the number of vertices, 
edges, and facets of the two polyhedra and $k$ is the (similarly defined) complexity
of their intersection;
(b) an output-sensitive algorithm for constructing an arrangement 
of $n$ triangles in 3-space in $O(n^{8/5+\eps} + k\log k)$ time, 
where $k$ is the output size;
and (c) an algorithm that, in $O(n^{8/5+\eps})$ expected time, 
counts all pairs of intersecting lines, in a set of $n$ lines in 3-space.

\paragraph{Background.}
Algorithmic problems that involve lines in three dimensions have been studied for more than 30 years,
covering the problems mentioned above and several others.
An early study by McKenna and O'Rourke~\cite{MO} has developed some of the tools and techniques
for tackling these problems.
Various techniques for 
ray shooting, and for the related problems of computing and verifying depth orders and hidden surface removal
have been studied in de Berg's dissertation~\cite{dB}, and later by de Berg et al.~\cite{BHOSK}.
Another work that developed some of the infrastructure for these problems is by 
Chazelle et al.~\cite{CEGSS}, who presented several combinatorial and algorithmic 
results for problems involving lines in 3-space. 
Agarwal and Matou\v{s}ek~\cite{AM} reduced ray shooting problems, via parametric search,
to segment emptiness problems (where the query is a segment and we want to determine whether
it intersects any input object), and obtained efficient solutions via this reduction.
See also~\cite{MS} and~\cite{AgS}
for studies of some additional and special cases of the ray shooting problem.

Most of the works cited above suffer from the `curse' of the four-dimensionality of 
(the parametric representation of) lines in space, which leads to algorithms whose complexity
is inferior to those obtained in our work. Nevertheless, there are a few instances where
better solutions can be obtained, such as in \cite{CEGS,CEGSS} and some other works.

\paragraph{Our results.}
Using the polynomial partitioning technique of \cite{Guth,GK},
we derive more efficient algorithms for the problems listed above.
In our first main result, presented in Section~\ref{sec:shoot},
we tackle the ray-shooting problem,
and construct a data structure on an input set of $n$ 
triangles, which requires $O(n^{3/2+\eps})$ storage and preprocessing,
so that a ray shooting query can be answered in $O(n^{1/2+\eps})$ time,
for any $\eps>0$. We then extend the technique, in Section~\ref{sec:seg},
to obtain an equally-efficient data structure for the segment-triangle 
intersection reporting, emptiness, and approximate counting problems, 
where in the case of approximate counting the query time bound has an
additional term that is nearly linear in the output size.

These are significant improvements over previous results, which, as already noted, 
have treated the lines supporting the edges of the input triangles and the 
line supporting the query ray (or segment) 
as points or surfaces in a suitable four-dimensional parametric space
(in many of the earlier works, lines were actually represented as points on the
\emph{Klein quadric} in five-dimensional projective space; see~\cite{BR,Hu,Pel:surv,St}). 
As a result, the algorithms obtained by these techniques were less efficient.

A weakness, or rather an intriguing peculiarity, of our analysis
is that it does not provide a desirably sharp tradeoff between storage 
and query time. To make this statement more precise, the tradeoff that
the earlier solutions provide, say for the ray shooting problem for
specificity, is that, for $n$ input triangles and with $s$ storage, 
for $s$ varying between $n$ and $n^4$, a ray-shooting query takes 
$O(n^{1+\eps}/s^{1/4})$ time; see, e.g.,~\cite{Pel:surv} 
(the `$4$' in the exponent comes from the fact that lines in 3-space are 
represented as objects in four-dimensional parametric space). Thus, with 
storage $O(n^{3/2+\eps})$, which is what our solution uses, the query time 
becomes about $O(n^{5/8})$, considerably weaker than our bound.

\begin{figure}[htb]
  \begin{center}
    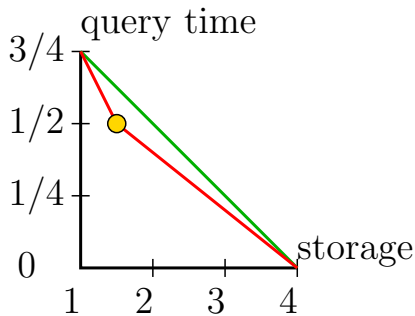
    \caption{{\sf The old tradeoff (green) and the new tradeoff (red). The $x$-axis is the 
    storage as a function of $n$, and the $y$-axis is the query cost. Both axes are drawn in a logarithmic scale.} }
    \label{tradeoff}
  \end{center}
\end{figure}

An ambitious, and maybe unrealistic goal would be to improve the tradeoff
so that the query time is only $O(n^{1+\eps}/s^{1/3})$. (This does indeed 
coincide with the bound that our main result gives, as the storage that 
it uses is $O(n^{3/2+\eps})$, but this coincidence only holds for this 
amount of storage.) Although not achieving this goal, still,
combining our technique with the known, aforementioned `$4$-dimensional' 
tradeoff, we are able to obtain an `in between' tradeoff, 
which we present in Section~\ref{sec:trade}. Concretely, the tradeoff
is that, with $s$ storage, the cost of a query is
\begin{equation}
  \label{eq:trade1}
  Q(n,s) = \begin{cases}
    O\left( \frac{n^{5/4+\eps}}{s^{1/2}} \right) , & s = O(n^{3/2+\eps}) , \\
    O\left( \frac{n^{4/5+\eps}}{s^{1/5}} \right) , & s = \Omega(n^{3/2+\eps}) .
  \end{cases}
\end{equation} 
Note that this tradeoff contains our bounds 
$(s,Q) = \left(O(n^{3/2+\eps}), O(n^{1/2+\eps})\right)$, as a special case, that
at the extreme ends $s=\Theta(n)$, $s=\Theta(n^4)$, of the range of $s$ we get
$Q = O(n^{3/4+\eps})$, $Q = O(n^\eps)$, respectively,\footnote{%
  The actual query time in the older tradeoff, with maximum storage, is $Q = O(\log n)$.}
as in the older tradeoff, and that the new tradeoff is better for any in-between value of $s$.
A comparison between the two tradeoffs is illustrated in Figure~\ref{tradeoff}.
Our improved tradeoff applies to all the problems studied in this paper.
In particular, it implies that, in all these problems, the overall cost of processing 
$m$ queries with $n$ input objects, including preprocessing cost, is
\begin{equation} \label{eq:trade2}
  \max\Bigl\{ O(m^{2/3}n^{5/6+\eps} + n^{1+\eps}),\; O(n^{2/3}m^{5/6+\eps} + m^{1+\eps}) \Bigr\} ,
\end{equation}
for any $\eps>0$;
for the output-sensitive problems, this bounds the total overhead cost.
The first (resp., second) bound dominates when $n\ge m$ (resp., $n\le m$).

We then present, in Section~\ref{sec:other}, extensions of our technique
for solving the other problems (iii), (iv) and (v) listed above.
In all these applications,
our algorithms are output-sensitive for the reporting versions,
so that the query time bound, or the full processing cost bound, contains an 
additional term that is nearly linear in the output size.
See Section~\ref{sec:other} for the concrete bounds that we obtain.

\section{Ray shooting amid triangles}
\label{sec:shoot}

Let $\T$ be a collection of $n$ triangles in ${\reals}^3$. We fix some sufficiently large 
constant parameter $D$, and construct a partitioning polynomial $f$ of degree 
$O(D)$ for $\T$, so that each of the $O(D^3)$ connected components $\tau$ 
of $\reals^3\setminus Z(f)$ (the cells of the partition) is crossed by 
at most $n/D^2$ triangle edges. We refer to triangles whose edge crosses $\tau$ 
as \emph{narrow triangles} (with respect to $\tau$),
and refer to the remaining triangles that cross $\tau$ (but none of their edges do) as
\emph{wide triangles}. We denote the set of narrow (resp., wide)
triangles in $\tau$ by $\N_\tau$ (resp., $\W_\tau$). The existence 
of such a partitioning polynomial is implied, as a special case, 
by the general machinery developed in Guth~\cite{Guth}.
An algorithm for constructing $f$ is given in a recent work of
Agarwal et al.~\cite{AAEZ}. It runs in $O(n)$ time, for any constant value 
of $D$, where the constant of proportionality depends (polynomially) on $D$.

For technical reasons, we want to turn any query ray into a bounded segment,
and we do it by enclosing all the triangles of $\T$ by a sufficiently large 
bounding box $B_0$, and by clipping any query ray to its portion within $B_0$. 

For each (bounded) cell $\tau\subseteq B_0$ of the partition, we take 
the set $\W_\tau$ of wide triangles in $\tau$,
and prepare a data structure for efficient segment-shooting queries into
the triangles of $\W_\tau$, by segments that are
fully contained in $\tau$. The nontrivial details of this procedure
are given in Section~\ref{sec:wide}. As we show there, we can
construct such a structure with storage and preprocessing
$O(|\W_\tau|^{3/2+\eps}) = O(n^{3/2+\eps})$, for any $\eps > 0$ 
(where the choice of $D$ depends on $\eps$), and each segment-shooting 
query takes $O(|\W_\tau|^{1/2+\eps}) = O(n^{1/2+\eps})$ time.

The preprocessing then recurses within each such cell $\tau$ of the partition, 
with the set $\N_\tau$ of the narrow triangles in $\tau$. The recursion 
terminates when the number of input triangles becomes smaller than the 
constant threshold $n_0 := O(D^2)$, in which case we simply output the 
list of triangles in the subproblem. 


A query, with a ray (now turned into a segment) $\rho$, emanating 
from a point $q$, is answered as follows. 
We first consider the case where $\rho$ (that is, the line containing $\rho$) is not
fully contained in $Z(f)$, and discuss the (simpler, albeit still involved) 
case where $\rho\subset Z(f)$, later.

\paragraph{The case where $\rho\not\subset Z(f)$.}
We assume a standard model of algebraic computation, in which a variety of computations
involving polynomials of constant degree, such as computing (some discrete representation of) 
the roots of such polynomials,
performing comparisons and algebraic computations (of constant degree) with these roots, 
and so on, can be done exactly in time $C(\delta)$, where $\delta$ is the degree of 
the polynomial, and $C(\delta)$ is a constant that depends on $\delta$; see, e.g., \cite{AEZ,BPR}.

\begin{figure}[htb]
  \begin{center}
    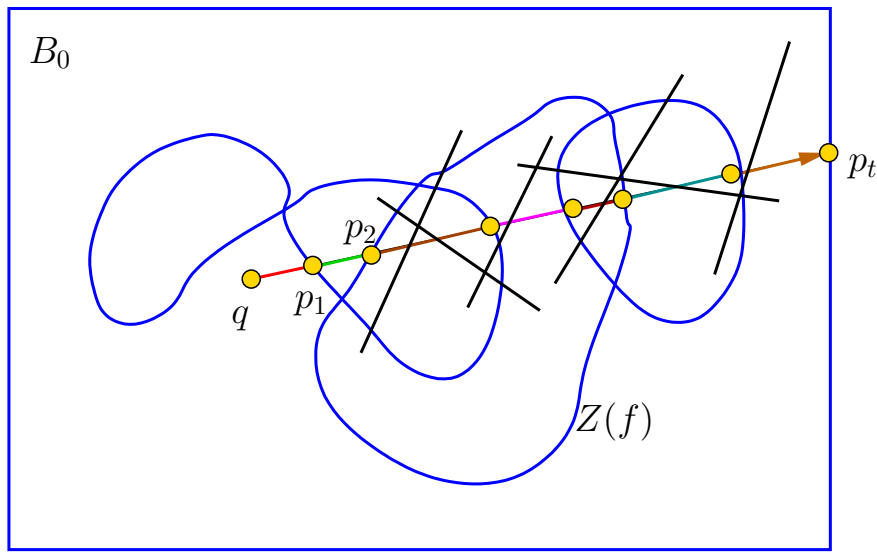
    \caption{{\sf A two-dimensional rendering of the the general structure of the ray-shooting mechanism.} }
    \label{cellbycell}
  \end{center}
\end{figure}

Using this model, we first locate the cell of the partition that 
contains the starting endpoint $q$ of the segment $\rho$, in constant time 
(recalling that $D$ is a constant). One way of doing this is to construct
the \emph{cylindrical algebraic decomposition (CAD)} of $Z(f)$ 
(see \cite{Col,SS2}), associate with each cell $\sigma$ of the CAD 
the cell of $\reals^3\setminus Z(f)$ that contains it (or an indication 
that $\sigma$ is contained in $Z(f)$), and then search with $q$ in
the CAD, coordinate by coordinate (see, e.g.,~\cite{AAEZ} for more details concerning
such an operation).
We then find, in constant time,
the $t=O(D)$ points of intersection of $\rho$ with $Z(f)$,
and sort them into a sequence $P := (p_1,\ldots,p_t)$ in their order 
along $\rho$; we assume that $p_t\in\bd B_0$, and ignore the suffix 
of $\rho$ from $p_t$ onwards.
The points in $P$ partition $\rho$ into a sequence of segments, each 
of which is a connected component of the intersection of $\rho$ with some
cell. The first segment is $e_1 = qp_1$, the subsequent segments are
$e_2 = p_1p_2$, $e_3 = p_2p_3,\ldots,e_t = p_{t-1}p_t$. We denote 
by $\tau_i$ the cell containing the $i$-th segment, for $i=1,\ldots,t$
(a cell can appear several times in this sequence). See Figure~\ref{cellbycell}.

We now process the segments $e_i$ in order. For each segment $e_i$, 
let $\tau_i$ denote the partition cell that contains $e_i$. We first
perform a ray-shooting (or rather a segment-shooting) query in the 
structure for $\W_{\tau_i}$ with the segment $e_i$. As already mentioned
(and will be described in Section~\ref{sec:wide}), this step can be performed in 
$O(n^{1/2+\eps})$ time, with $O(n^{3/2+\eps})$ storage and 
preprocessing, for any $\eps > 0$. We then query with $e_i$ in 
the substructure recursively constructed for $\N_{\tau_i}$. If at least
one of the two queries succeeds, i.e., outputs a point that lies 
on $e_i$, we report the point nearest to the starting point of $e_i$, 
and terminate the whole query. If both queries fail, we proceed to the 
next segment $e_{i+1}$ and repeat this step. If the mechanism fails for all 
the segments, we report that $\rho$ does not hit any triangle of $\T$.

\paragraph{The case where $\rho\subset Z(f)$.}
We use the cylindrical algebraic decomposition (CAD) of $Z(f)$ (see \cite{Col,SS2}), 
which has already been constructed for the earlier case. One of its by-products
is a \emph{stratification} of $Z(f)$, which is a decomposition of $Z(f)$ 
into pairwise disjoint relatively open patches of dimensions $0$, $1$, 
and $2$, called \emph{strata} (each stratum is a cell of the CAD),
so that each of the two-dimensional strata is $xy$-monotone and its 
relative interior is free of any singularities of $Z(f)$, and $Z(f)$ 
is the union of the closures of these two-dimensional strata, excluding 
possible components of $Z(f)$ of dimension at most $1$, which we may ignore.
We compute the intersection arcs $\gamma_\Delta := Z(f)\cap\Delta$, for 
$\Delta\in\T$, and distribute each arc amid the closures of the
two-dimensional strata that it traverses. 
We then project the closure of each two-dimensional stratum $\sigma$ onto the $xy$-plane, 
including the portions of the arcs $\gamma_\Delta$ that the closure contains, 
and preprocess the resulting collection of $O(n)$ algebraic arcs, each of 
degree $O(D) = O(1)$, into a planar ray-shooting data structure, whose
details are spelled out in Section~\ref{sec:onzf}.\footnote{%
  This specific planar ray-shooting problem, amid constant-degree algebraic arcs,
  has not received full attention in the past, although several algorithms have
  been proposed, mostly with suboptimal solutions. Consult, e.g., 
  Table 2 in Agarwal~\cite{Ag:rs}; see also \cite{AvKO-93,Kol}.}
As we show there, we can 
answer a ray-shooting query in $O(n^{1/2+\eps})$ time, using $O(n^{3/2+\eps})$ storage,
for any $\eps > 0$, where the constants of proportionality depend on $\eps$, as does
the choice of $D$. The overall storage complexity, over all the (projected) strata 
of $Z(f)$, is thus $O(n^{3/2+\eps})$, and the overall query time, over all strata 
met by the query ray $\rho$, is $O(n^{1/2+\eps})$, for a larger constant of 
proportionality (that depends on $\eps$).

Note that the recursion on $D$ when the query ray comes to lie on the zero set
of the current partitioning polynomial. When this happens, we solve the problem 
in this recursive instance using the (nonrecursive) procedure in Section~\ref{sec:onzf}
and terminate the (current branch of the) recursion. Another way of saying this is that 
the leaves of the $D$-recursion tree represent either constant-size subproblems
or subproblems on the zero set of the current partitioning polynomial, and the 
inner nodes represent subproblems of shooting within the partition cells. 

\paragraph{Analysis.}
The correctness of the procedure is fairly easy to establish.
Denote by $S(n)$ the maximum storage used by the structure
for a set of at most $n$ triangles, and denote by $S_0(n)$ (resp., $S_1(n)$)
the maximum storage used by the auxiliary structure for a set of at most $n$ wide 
triangles in a cell of the partition, as analyzed in Section~\ref{sec:wide}
(resp., for a set of at most $n$ intersection arcs on $Z(f)$, which we process 
for planar ray-shooting in Section~\ref{sec:onzf}). Then $S(n)$ obeys the recurrence
\begin{equation}
  \label{eq:storage}
  S(n) = O(D^3)S_0(n) + S_1(n) + O(D^3)S(n/D^2) , 
\end{equation}
for $n > n_0$, and $S(n) = O(n)$ for $n\le n_0$, where 
$n_0 := c D^2$, for a suitable constant $c \ge 1$. We show, in the respective
Sections~\ref{sec:wide} and~\ref{sec:onzf}, that $S_0(n) = O(n^{3/2+\eps})$ 
and $S_1(n) = O(n^{3/2+\eps})$, for any $\eps > 0$, 
where both constants of proportionality depend on $D$ and $\eps$,
from which one can easily show that the solution of (\ref{eq:storage}) is
$S(n) = O(n^{3/2+\eps})$, for a slightly larger, but still arbitrarily small
$\eps>0$; to achieve this bound, we need to take $D$ to be $2^{\Theta(1/\eps)}$, 
as will follow from our analysis.  Regarding the bound on the preprocessing 
time $T(n)$, we obtain a similar recurrence as in~(\ref{eq:storage}), namely, 
\[
T(n) = O(n) + O(D^3)T_0(n) + T_1(n) + O(D^3)T(n/D^2) , 
\]
where the non-recursive linear term is the time
to compute the polynomial $f$, and $T_0(n)$, $T_1(n)$ are defined in an analogous manner as above, and have similar upper
bounds as $S_0(n)$, $S_1(n)$ (see Sections~\ref{sec:wide} and~\ref{sec:onzf}).

Similarly, denote by $Q(n)$ the maximum 
query time for a set of at most $n$ triangles, and denote by $Q_0(n)$ (resp., $Q_1(n)$)
the maximum 
query time in the auxiliary structure for a set of at most $n$ wide triangles in a cell of the partition
(resp., for a set of at most $n$ intersection arcs within $Z(f)$, 
when the query ray lies on $Z(f)$). Then $Q(n)$ obeys the recurrence
\begin{equation}
  \label{eq:query}
  Q(n) = \max\left\{ O(D)Q_0(n) + O(D)Q(n/D^2) ,\; Q_1(n) \right\} ,
\end{equation}
for $n > n_0$, and $Q(n) = O(n) = O(1)$ for $n\le n_0$. 
(This reflects the observation, made above, that the current branch
of the recursion terminates when the query ray lies on the zero set of
the current partitioning polynomial.)
Again, the analysis in Sections~\ref{sec:wide} and~\ref{sec:onzf} 
shows that $Q_0(n) = Q_1(n) = O(n^{1/2+\eps})$, for any $\eps > 0$ 
(where the choice of $D$ depends on $\eps$, as above), from which one can 
easily show, using induction on $n$, that the solution of (\ref{eq:query}) is
$Q(n) = O(n^{1/2+\eps})$, for a slightly larger but still arbitrarily small $\eps>0$.

The main result of this section is therefore:
\begin{theorem}
  \label{thm:trimain}
  Given a collection of $n$ triangles in three dimensions, and
  a prescribed parameter $\eps>0$, we can process the triangles into a data structure
  of size $O(n^{3/2+\eps})$, in time $O(n^{3/2+\eps})$, so that a ray shooting query
  amid these triangles can be answered in $O(n^{1/2+\eps})$ time.
\end{theorem}

\subsection{Ray shooting into wide triangles} 
\label{sec:wide}

\paragraph{Preliminaries.}
In this subsection we present and analyze our procedure for ray shooting 
in the set $\W_\tau$ of the wide triangles in a cell $\tau$ of the partition.
We then present, in Section~\ref{sec:onzf}, a different approach that yields 
a procedure for ray shooting within $Z(f)$. Both procedures have the performance
bounds stated in Theorem~\ref{thm:trimain}. The efficiency of our structures 
depends on $D$ being a constant, since the constants of proportionality
depend polynomially (and rather poorly) on $D$. 

\begin{figure}[htb]
  \begin{center}
    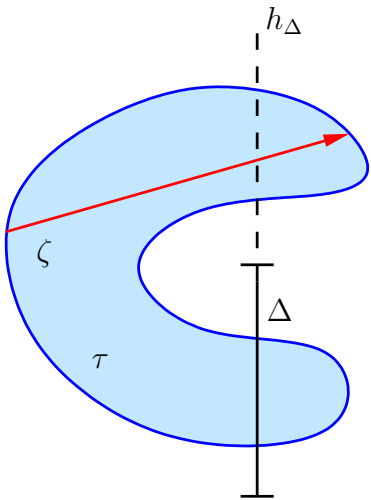
    \caption{{\sf Wide triangles cannot be replaced by their supporting planes
      for ray shooting within $\tau$.} }
    \label{fig:horseshoe}
  \end{center}
\end{figure}

We thus focus now on ray shooting in a set of wide triangles within a 
three-dimensional cell $\tau$ of the partition.
To appreciate the difficulty in solving this subproblem, we make the
following observation. A simple-minded approach might be to
replace each wide triangle $\Delta\in\W_\tau$ by the plane $h_\Delta$
supporting it. Denoting the set of these planes as $\H_\tau$, we
could then preprocess $\H_\tau$ for ray-shooting queries, each of which
specifies a query ray $\zeta$ and asks for the first intersection
of $\zeta$ with the planes of $\H_\tau$. Using standard machinery (see, e.g.~\cite{Ag:rs}),
this would result in an algorithm with the performance bounds that we want.
However, this approach is problematic, since, even though $\Delta$ is wide in $\tau$,
$h_\Delta$ could intersect $\tau$ in several connected components,
some of which lie outside $\Delta$. See Figure~\ref{fig:horseshoe}
for an illustration. In such cases, ray shooting amid the planes in 
$\H_\tau$ is not equivalent to ray shooting amid the triangles of 
$\W_\tau$, even for rays, or rather portions thereof, that are contained in $\tau$.

Our solution is therefore more involved, and proceeds as follows.

\paragraph{Canonical sets of wide triangles.}

Consider first, for exposition sake, the case where the starting point
of the shooting segment lies on $\bd\tau$ (the terminal point always 
lies on $\bd\tau$). As we will show, for each such segment query, the 
set of wide triangles in $\W_\tau$ that it intersects can be decomposed 
into a small collection of precomputed ``canonical'' subsets, where in each canonical set 
the wide triangles can be treated as planes (for that particular query segment). 
We show below that the overall size of these sets, over all possible segment 
queries, is $O(n^{3/2+\eps})$, for any $\eps > 0$. 

Actually, to prepare for the complementary case, where the
starting point of the query segment lies inside $\tau$, we calibrate our
algorithm, so that we control the storage that it uses, and consequently
also the query time bound. To do so, we introduce a \emph{storage parameter} $s$,
which can range between $n$ and $n^2$, as a second input to our procedure,
and then require that the actual storage and preprocessing cost be both 
$O(s^{1+\eps})$, for any $\eps > 0$. This relaxed notion of storage offers 
some simplification in the analysis. (We will also allow larger values of $s$
when we discuss tradeoff between storage and query time, in Section~\ref{sec:trade}.)

For each $\Delta\in\W_\tau$, let $\gamma_\Delta$ denote the intersection
curve of $\Delta$ with $\bd\tau$. Note that $\gamma_\Delta$ does not
have to be connected---it can have up to $O(D^2)$ connected components,
by Harnack's curve theorem~\cite{Harnack} (applied on the plane containing 
$\Delta$). Note also that $\bd\tau$ does not have to be connected, so 
$\gamma_\Delta$ can have nonempty components on different connected components 
of $\bd\tau$, as well as several components on the same connected component of $\bd\tau$.

We construct the locus $S_\tau$ of points on $\bd\tau$ that are 
either singular points of $Z(f)$ or points with $z$-vertical tangency.
Since $D$ is constant, $S_\tau$ is a curve of constant degree
(by B\'ezout's theorem, its degree is $O(D^2)$). 
We take a random sample $\R$ of $r_0$ triangles of $\W_\tau$, where the
analysis dictates that we choose $r_0 = D^{\Theta(1/\eps)}$, for the 
arbitrarily small prescribed $\eps>0$. Since we have chosen $D$ to be 
$2^{\Theta(1/\eps)}$, the actual choice of $r_0$ is $2^{\Theta(1/\eps^2)}$.

Let $\Gamma_\R = \{\gamma_\Delta \mid \Delta\in\R\}$, 
and let $\A_0 = \A(\Gamma_\R\cup\{S_\tau\})$ denote the arrangement 
of these curves within $\bd\tau$, together with $S_\tau$. 
By construction, each face of $\A_0$ is $xy$-monotone and does not 
cross any other branch of $Z(f)$ (at a singular point).
We partition each face $\varphi$ of $\A_0$ into pseudo-trapezoids (called trapezoids for short), 
using a suitably adapted version of a two-dimensional vertical decomposition scheme.
Let $\A_0^*$ denote the collection of these trapezoids on $\bd\tau$.
The number of trapezoids in $\A_0^*$ is proportional to the complexity
of $\A_0$, which is $O_D(r_0^2) = O(1)$ (we use the notation $O_D(\cdot)$
to indicate that the constant of proportionality depends on $D$, and
recall that $r_0$ also depends on $D$).

We assume that the trapezoids are relatively open. 
To cover all possible cases, we also include in the collection of
trapezoids the relatively open subarcs of arcs in $\Gamma_\R$
that the partition generates, the vertical edges of the trapezoids, 
and the vertices of the partition, but, for exposition sake, we will
only handle here the case of two-dimensional trapezoids. (The inclusion
of lower-dimensional `trapezoids' is simpler to handle; it does not affect 
the essence of the forthcoming analysis, nor does it affect the asymptotic
performance bounds.) 

Let $\psi_1$, $\psi_2$ be two distinct trapezoids of $\A_0^*$. 
Let $S(\psi_1,\psi_2)$ denote the collection of all segments
$e$ such that one endpoint of $e$ lies in $\psi_1$, the other 
endpoint lies in $\psi_2$, and the relative interior of $e$ 
is fully contained in the open cell $\tau$. We can parameterize
such a segment $e$ by four real parameters, so that two parameters 
specify the starting endpoint of $e$ (as a point in $\psi_1$, using,
e.g., the $xy$-parameterization of the $xy$-monotone face containing $\psi_1$),
and the other two parameters similarly specify the other endpoint.
(Fewer parameters are needed when lower-dimensional trapezoids are involved.)
Denote by $\F$ the corresponding (at most) four-dimensional parametric space.
Since each of $\tau$, $\psi_1$, $\psi_2$ is of constant complexity,
$S(\psi_1,\psi_2)$ is a semi-algebraic set in $\F$ of constant complexity. 
More specifically, we can write $S(\psi_1, \psi_2)$ 
as an (implicitly) quantified formula of the form
\[
S(\psi_1, \psi_2) = \{ (p_1, p_2) \mid p_1 \in \psi_1, p_2 \in \psi_2, \; \mbox{and} \; p_1 p_2 \subset \tau \},
\]
where $p_1 p_2$ denotes the line-segment connecting $p_1$ to $p_2$. 
Using the singly exponential quantifier-elimination algorithm 
in~\cite[Theorem 14.16]{BPR}, we can construct, in $O_D(1)$ time, a 
quantifier-free semi-algebraic representation of $S(\psi_1,\psi_2)$ 
of $O_D(1)$ complexity. Moreover, we can decompose $S(\psi_1, \psi_2)$ 
into its connected components, in $O_D(1)$ time as well.

For each segment $e \in S(\psi_1,\psi_2)$, let $\T(e)$ denote the
set of all wide triangles of $\W_\tau$ that $e$ crosses. We have
the following technical lemma.
\begin{lemma}
  \label{lem:cross}
  In the above notations, each connected component $C$ of
  $S(\psi_1,\psi_2)$ can be associated with a fixed set $\T_C$ of
  wide triangles of $\W_\tau$, none of which crosses $\psi_1\cup\psi_2$, 
  so that, for each segment $e\in C$, $\T_C\subseteq \T(e)$,
  and each triangle in $\T(e)\setminus\T_C$ crosses either $\psi_1$
  or $\psi_2$.
\end{lemma}

\begin{figure}[htb]
  \begin{center}
    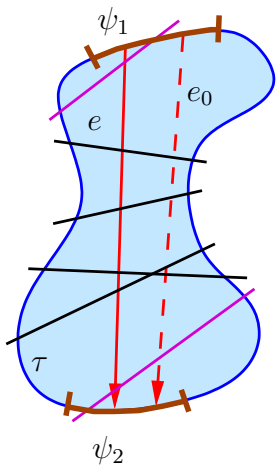
    \caption{{\sf The set $\T_C$ (consisting of the triangles depicted as black segments),
      and an illustration of the proof of Lemma~\ref{lem:cross}.}}
    \label{fig:tcee0}
  \end{center}
\end{figure}

\medskip
\noindent{\bf Proof.}
Pick an arbitrary but fixed segment $e_0$ in $C$, and define $\T_C$ to consist
of all the triangles in $\T(e_0)$ that do not cross $\psi_1\cup\psi_2$.
See Figure~\ref{fig:tcee0} for an illustration.

Let $e$ be another segment in $C$. Since $C$ is connected, as a set in
$\F$ (recall that this is a four-dimensional parametric space representing the segments),
there exists a continuous path $\pi$ in $C$ that connects $e_0$ and $e$
(recall that each point on $\pi$ represents a segment with one endpoint on
$\psi_1$ and the other on $\psi_2$, and $\pi$ represents a continuous variation
of such a segment from $e_0$ to $e$).
Let $\Delta$ be a triangle in $\T(e_0)$ that does not cross 
$\psi_1\cup\psi_2$ (that is, $\Delta\in\T_C$), and let $h_\Delta$ denote its supporting plane.
As a segment $e'$ traverses $\pi$ from $e_0$ to $e$, the point $q_\Delta(e') := e'\cap h_\Delta$ 
is well defined and varies continuously in $\tau$, unless $e'$ comes to be contained in, 
or parallel to $h_\Delta$, a situation that, as we now argue, cannot arise.

In what follows, we are going to argue that the segment $e'$ is detached from $\Delta$ when either
(i) the relative interior of $e'$ touches the boundary of $\Delta \cap \tau$,
which cannot happen since then $e'$ would have to (partially) exit $\tau$ and meet its boundary,
contrary to the assumption that $e'$ is fully contained in $\tau$ (recall that $\tau$ is open), or
(ii) $\Delta \cap \tau$ touches an endpoint of $e'$, which again cannot happen because
the endpoints of $e'$ lie on $\psi_1\cup\psi_2$, and $\Delta$ is assumed not to intersect $\psi_1\cup\psi_2$.
More formally, we argue as follows. By assumption, $q_\Delta(e_0)$ lies in $\Delta$, and, 
as long as $q_\Delta(e')$ is defined (i.e., $e'$ intersects $\Delta$ at a unique point), 
$q_\Delta(e')$ cannot reach $\bd\Delta$ because the corresponding 
segment $e'$ is fully contained in the open cell $\tau$ and $\Delta$ 
is wide in $\tau$. (We may assume that $e'$ does not overlap $\Delta$,
because, since $\Delta$ is wide, that would mean that both endpoints 
of $e'$ lie on $\Delta$, but then $\Delta$ crosses both $\psi_1$ and
$\psi_2$, which we have assumed not to be the case.) We claim that $q_\Delta(e')$ 
must be nonempty throughout the motion of $e'$ along $\pi$, 
for otherwise $q_\Delta(e')$ would have to reach an endpoint 
of $e'$, which, by definition of $S(\psi_1,\psi_2)$, 
must lie on $\psi_1$ or on $\psi_2$. But then $\Delta$ would have 
to intersect either $\psi_1$ or $\psi_2$, contrary to assumption.
It follows that $q_\Delta(e)$ also lies in $\Delta$, so $\Delta\in\T(e)$.
This establishes the first assertion of the lemma. 

We next need to show that each triangle in $\T(e)\setminus\T_C$ must 
cross either $\psi_1$ or $\psi_2$, which is our second assertion.
Let $\Delta$ be a triangle in $\T(e)\setminus\T_C$, and assume to the 
contrary that $\Delta$ does not cross $\psi_1\cup\psi_2$. We run the 
preceding argument in reverse (moving from $e$ to $e_0$), and observe 
that, by assumption and by the same argument (and notations) as above, 
$q_\Delta(e')$ remains inside $e'$, for all intermediate segments $e'$ 
along the connecting path $\pi$, and does not reach $\bd{\Delta \cap \tau}$,
so $\Delta\in \T(e_0)$ and thus also $\Delta\in \T_C$ (by definition
of $\T_C$), contradicting our assumption. 
This establishes the second assertion, and thereby completes the proof.
$\Box$

\medskip
Lemma~\ref{lem:cross} and its proof show that, for each connected component 
$C$ of $S(\psi_1,\psi_2)$, the canonical set $\T_C$, of wide triangles 
that are crossed by all segments in $C$ and do not cross $\psi_1\cup\psi_2$,
assigned to $C$, is unique and is independent of the choice of $e_0$.
(This is because the sets $\T(e_0)$, for $e_0\in C$, differ from each other
only in triangles that cross either $\psi_1$ or $\psi_2$.)
The collection of all these sets $\T_C$, over all connected components 
$C$, and all pairs of trapezoids $(\psi_1,\psi_2)$, is part of the whole output
collection of canonical sets over $\tau$; the rest of this collection
is constructed recursively over the trapezoids $\psi$ of $\A_0^*$.

\paragraph{The algorithm.}

For each trapezoid $\psi$ of $\A_0^*$, the \emph{conflict list}
$K_\psi$ of $\psi$ is the set of all wide triangles that cross $\psi$. 
By standard random sampling arguments~\cite{CEGS-91}, with high probability, 
the size of each conflict list is $O\left(\frac{n}{r_0}\log r_0\right)$,
where the constant of proportionality depends on $D$.

Two extreme situations that require special treatment are (i)
$0$-dimensional trapezoids (vertices), where there is no bound on the 
number of triangles of $\W_\tau$ that can contain a vertex $\psi$
(in which case we do not recurse at $\psi$), 
but it suffices just to maintain one of them in the structure,
because if the starting (or the other) endpoint of a query segment lies
at $\psi$, it does not matter which of these incident triangles we use.
Also, we do not recurse at $\psi$ (technically, it has no conflict list, as
no triangle crosses it).
(ii) triangles that fully contain a two-dimensional trapezoid $\psi$,
where these triangles are contained in some planar component of $Z(f)$
(where only the triangles that cross $\psi$ are processed recursively).
We assume for simplicity that these triangles do not overlap one another.
Since we are handling here rays that are not contained in $Z(f)$,
such a ray $\rho$ can cross $Z(f)$ at only $O(D) = O(1)$ points, and
it is easy to find, in $O(1)$ time, these crossing points, and then check,
in $O(\log n)$ time (with linear storage), whether any of these points 
belongs to a triangle contained in $Z(f)$. (The case where the triangles
are overlapping is also easy to handle. The performance bounds deteriorate, 
but are still within the overall bounds that we derive.)

For every pair of trapezoids $\psi_1$, $\psi_2$, we compute 
$S(\psi_1,\psi_2)$ and decompose it into its connected components.
We pick some arbitrary but fixed segment $e_0$ from 
each component $C$, compute the set $\T(e_0)$ of the wide triangles 
that cross $e_0$, and remove from it any triangle that 
crosses $\psi_1\cup\psi_2$, thereby obtaining the set $\T_C$.
All this takes $O_D(r_0^4 n) = O_D(n)$ time,
and the overall size of the produced canonical sets is also $O_D(n)$.

Let $s$ be the storage parameter associated with the problem,
as defined earlier, and recall that we require that $n\le s \le n^2$.
Each canonical set $\T_C$ is preprocessed into a data structure that 
supports ray shooting in the set of \emph{planes}
$\H_C = \{h_\Delta \mid \Delta\in\T_c\}$, where $h_\Delta$ is
the plane supporting $\Delta$. We construct these structures 
so that they use $O(s^{1+\eps})$ storage (and preprocessing), for any $\eps>0$,
and a query takes $O(n\;{\rm polylog}(n)/s^{1/3})$ time (see, e.g.,~\cite{Ag:rs}). 

We now process recursively each conflict list $K_\psi$, over all
trapezoids $\psi$ of $\A_0^*$. Each recursive subproblem uses
the same parameter $r_0$, but now the storage parameter that we 
allocate to each subproblem is only $s/r_0^2$. We keep recursing 
until we reach conflict lists of size close to $n^2/s$.
More precisely, after $j$ levels of recursion, we get a total of at most 
$(c_0 r_0^2)^j = c_0^jr_0^{2j}$ subproblems, each involving at most
$\left(\frac{c_1\log r_0}{r_0}\right)^j n$ wide triangles, for some 
constants $c_0$, $c_1$ that depend on $D$, and thus on $\eps$ (specifically, 
$c_1$ depends on $c_0$ and $c_0 = O(D^2)$ by B\'ezout's theorem),
but are considerably smaller than $r_0$, which, as already mentioned, 
we take to be $D^{\Theta(1/\eps)}$. 

We stop the recursion at the first level $j^*$ at which
$(c_1r_0\log r_0)^{j^*} \ge s/n$.
As a result, we have ${r_0}^{j^{*}} \le s/n$, and we get
$c_0^{j^{*}} r_0^{2j^{*}} = O(s^2/n^{2-\eps})$ subproblems, 
for any $\eps > 0$, where the choice of $D$ (and therefore also of $c_0$, $c_1$ 
and $r_0$) depends, as above, on $\eps$.  Each of these subproblems involves at most 
\[
\left(\frac{c_1\log r_0}{r_0}\right)^{j^*} n =
\left(\frac{(c_1\log r_0)^2}{c_1r_0\log r_0 }\right)^{j^*} n \le
(c_1\log r_0)^{2j^*}\cdot \frac{n^2}{s} = 
\frac{n^{2+\eps}}{s}  
\]
triangles, for any $\eps>0$. For this estimate to hold, we choose
$D = 2^{\Theta(1/\eps)}$.
Hence the overall size of the inputs, as well as of the canonical sets,
at all the subproblems throughout the recursion, is
${\displaystyle O\left(\frac{s^2}{n^{2-\eps}}\right) \cdot \frac{n^{2+\eps}}{s} = O(sn^{2\eps})} = O(s^{1+\eps})$,
for a slightly larger $\eps > 0$.

Note that the canonical sets that we encounter 
when querying with a fixed segment $e$
are not necessarily pairwise disjoint. This is because the sets 
$K_{\psi_1}$ and $K_{\psi_2}$ are not necessarily disjoint (they are
disjoint of $\T_C$, though). This does not pose a problem for 
ray shooting queries, but will be problematic for \emph{counting queries}; 
see Section~\ref{sec:seg}.

At the bottom of the recursion, each subproblem contains at most $n^{2+\eps}/s$ wide triangles,
which we merely store in the structure. As just calculated, the overall storage that
this requires is $O(s^{1+\eps})$, for a slightly larger $\eps$, as above.
We obtain the following recurrence 
for the overall storage $S(N_W,s_W)$ for the structure constructed on $N_W$ wide triangles,
where $s_W$ denotes the storage parameter allocated to the structure
(at the root  $N_W = n$, $s_W = s$).
\[
S(N_W,s_W) = \left\{
\begin{array}{ll}
  O_D(r_0^4 s_W^{1+\eps}) +
    c_0r_0^2 S\left(\frac{c_1N_W \log r_0}{r_0},\; \frac{s_W}{r_0^2}\right) & \mbox{for $N_W \ge n^{2+\eps}/s$,}\\[1mm]
  O(N_W) &\mbox{for $N_W < n^{2+\eps}/s$}.
\end{array}
\right\}
\]
(The overhead term is actually $O_D(r_0^4 N_W + r_0^4 s_W^{1+\eps})$, but the second term dominates.)
Throughout the recursion we have $N_W \le s_W \le N_W^2$.
Indeed, starting with $n$ and $s$, after $j$ recursive
levels we have $N_W \le \left(\frac{c_1\log r_0}{r_0}\right)^{j} n$ and $s_W = s/r_0^{2j}$.
Hence the right inequality continues to hold (for $s \le n^2$), and the
left inequality holds as long as $\left(\frac{c_1\log r_0}{r_0}\right)^{j} n \le s/r_0^{2j}$,
or $(c_1 r_0\log r_0)^{j} \le s/n$, which indeed holds up to the terminal level $j^*$, by construction.

Unfolding the first recurrence up to the terminal level $j^*$, where $N_W < n^{2+\eps}/s$, 
the sum of the nonrecursive overhead terms $O_D(r_0^4 s_W^{1+\eps})$, over all nodes at a fixed level $j$, is
\[
c_0^{j} r_0^{2j} \cdot O\left( \frac{s_W^{1+\eps}}{r_0^{2j(1+\eps)}} \right) =
O\left(\frac{c_0^j}{r_0^{2j\eps}} s_W^{1+\eps}\right) = O\left( s_W^{1+\eps} \right) ,
\]
by the choice of $r_0$. Hence, starting the recurrence at $(N_W,s_w) = (n,s)$, 
the overall contribution of the overhead terms (over the logarithmically many 
levels) is $O(s^{1+\eps})$, for a slightly larger $\eps$.
At the bottom of recurrence, we have, as already noted, 
$O(s^2/n^{2-\eps})$ subproblems, each with at most $O(n^{2+\eps}/s)$ triangles,
so the sum of the terms $N_W$ at the bottom of recurrence is also $O(s^{1+\eps})$.
In other words, the overall storage used by the data structure is $O(s^{1+\eps})$.
Using similar considerations, one can show that the overall preprocessing time is 
$O(s^{1+\eps})$ as well, since the time obeys essentially the same recurrence.

\paragraph{Answering a query.}

\begin{figure}[htb]
\begin{center}
  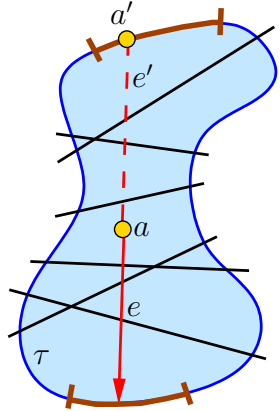
  \caption{\sf{Segment shooting from inside the cell $\tau$:
      Extending the segment backwards and the resulting canonical set of triangles.} }
  \label{inside}
\end{center}
\end{figure}

To perform a query with a segment $e$ that starts at a point $a$ (that 
lies anywhere inside $\tau$), we extend $e$ from $a$
backwards, find the first intersection point $a'$ of the resulting 
backward ray with $\bd\tau$, and denote by $e'$ the segment that starts 
at $a'$ and contains $e$. See Figure~\ref{inside} for an illustration.
This takes $O_D(1)$ time. This step is vacuous when $e$ starts on $\bd\tau$,
in which case we have $e' = e$.

We find the pair of trapezoids $\psi_1$, $\psi_2$ that contain the 
endpoints of $e'$, find the connected component $C\subseteq S(\psi_1,\psi_2)$
that contains $e'$, and retrieve the canonical set $\T_C$.
We then perform segment shooting along $e$ from $a$ in the structure 
constructed for $\H_C$, and then continue recursively in the subproblems 
for $K_{\psi_1}$ and $K_{\psi_2}$. We output the triangle that $e$ hits 
at a point nearest to $a$, or, if no such point is produced, report that 
$e$ does not hit any wide triangle inside $\tau$. In case both endpoints 
of $e'$ lie in the same trapezoid $\psi$ (that is, $\psi_1 = \psi_2$), we set 
$\T_C$ to be empty at this step (it is easy to verify that this indeed 
must be the case), and then continue processing $e'$ (and 
thus $e$) in the recursion on $K_\psi$.

The correctness of the procedure follows from the fact that $e'$ intersects
all the triangles of $\T_C$, and thus replacing these triangles by their
supporting planes cannot produce any new (false) intersection of any of these
triangles with $e$, and any other wide triangle that $e$ hits must belong
to $K_{\psi_1}\cup K_{\psi_2}$.

The query time $Q(N_W,s_W)$ satisfies the recurrence
\[
Q(N_W,s_W) = \left\{
\begin{array}{ll}
  O_D(1) + O\left(\frac{N_W {\rm polylog}(N_W)} {s_W^{1/3}} \right) 
+ 2Q\left(\frac{c_1 N_W \log r_0}{r_0},\; \frac{s_W}{r_0^2} \right) & \mbox{for $N_W \ge n^{2+\eps}/s$,}\\[1mm]
  O(N_W) & \mbox{for $N_W < n^{2+\eps}/s$} .
\end{array}
\right\}
\]
Unfolding the first recurrence, we see that when we pass from some 
recursive level to the next one, we get two descendant subproblems
from each recursive instance, and the term $N_W {\rm polylog}(N_W)/s_W^{1/3}$ 
is replaced in each of them by the (upper bound) term
\[
\frac{ \frac{c_1 N_W \log r_0}{r_0} }{ \left(\frac{s_W}{r_0^2} \right)^{1/3} } \cdot {\rm polylog}(N_W) = 
\frac{c_1 \log r_0}{r_0^{1/3}} \cdot \frac{N_W {\rm polylog}(N_W) }{s_W^{1/3}} .
\]
Hence the overall bound for the nonrecursive overhead terms in the 
unfolding, starting from $(N_W,s_W) = (n,s)$, is at most
\[
O\left( \sum_{j\ge 0}
\left( \frac{2c_1 \log r_0}{r_0^{1/3}} \right)^{j} \right) \cdot \frac{n \ {\rm polylog}(n)}{s^{1/3}} =
O\left( \frac{n \ {\rm polylog}(n)}{s^{1/3}} \right) ,
\]
provided that $r_0$ is sufficiently large. Adding the cost at the $2^{j^*}$ subproblems
at the bottom level $j^*$ of the recursion, where the cost of each subproblem is at 
most $n^{2+\eps}/s$, gives an overall bound for the query time of
\begin{equation} \label{eq:qns}
Q(n,s) = O\left( \frac{n \ {\rm polylog}(n) }{s^{1/3}} + \frac{n^{2+\eps}}{s} \right) .
\end{equation}
Starting with $s=n^{3/2}$, the query time is $O(n^{1/2+\eps})$.
We thus obtain 
\begin{proposition} \label{prop:inside}
  For a (bounded) cell $\tau$ of the polynomial partition, and a set $\W$ of 
  $n$ wide triangles in $\tau$, one can construct a data structure of size and 
  preprocessing cost $O(n^{3/2+\eps})$, so that a segment-shooting query within $\tau$, 
  from any starting point, can be answered in $O(n^{1/2+\eps})$ time, for any $\eps > 0$.
\end{proposition}
The case where the query ray is contained in $Z(f)$ is discussed in detail in 
the following Section~\ref{sec:onzf}, culminating in Proposition~\ref{prop:onzf},
with the same performance bounds. Thus, combined with the results in this 
subsection, Theorem~\ref{thm:trimain} follows.

\subsection{Ray shooting within $Z(f)$}
\label{sec:onzf}

We now consider the case where the (line supporting the) query ray is contained
in the zero set $Z(f)$ of $f$. We present our result in a more general form,
in which we are given a collection $\Gamma$ of $n$ constant-degree algebraic arcs
in the plane,\footnote{%
  Recall that we project each stratum of $Z(f)$ onto the $xy$-plane.}
and preprocess it into a data structure of $O(n^{3/2+\eps})$ size, which 
can be constructed in $O(n^{3/2+\eps})$ time, that supports ray-shooting queries 
in time $O(n^{1/2+\eps})$ per query, for any $\eps > 0$.

\paragraph{Ray shooting amid arcs in the plane.}
Let $\Gamma$ be a set of $n$ algebraic arcs of constant degree in the plane.
We may assume, after breaking each arc into a constant number of subarcs,
if needed, that each arc is $x$-monotone, has a smooth relative interior, and is either convex or concave. 
For concreteness, and with no loss of generality, we assume in what 
follows that all the arcs of $\Gamma$ are convex. That is, the tangent 
directions turn counterclockwise along each arc as we traverse it from left to right.

We present the solution in four steps. We first discuss the problem of detecting an
intersection between a query line and the input arcs. We then extend this machinery to
detecting intersection with a query ray, and finally to detecting intersection with 
a query segment. Once we have such a procedure, we can use the parametric-search
technique of Agarwal and Matou\v{s}ek~\cite{AM} (this is our fourth step) to perform 
ray shooting, with similar performance bounds. The reason for this gradual presentation 
of the technique is that each step uses the structure from the previous step as a
substructure.

We remind the reader, again, that the problem of ray shooting in the plane amid
a general collection of constant-degree algebraic arcs, which is the problem
considered in this section, does not seem to have a solution with sharp performance bounds;
see Table 2 in \cite{Ag:rs} and~\cite{AvKO-93,Kol}.

\subsubsection{Detecting line intersection with the arcs}
\label{sec:line}

Our approach is to transform the line-intersection problem to a planar point-location problem,
by mapping the lines to points and the arcs $\gamma \in \Gamma$ to semi-algebraic sets (whose
complexity depends on the complexity of $\gamma$). Our mapping is based on \emph{quantifier elimination},
and proceeds as follows.\footnote{%
  There is another, more direct approach to solving this problem, which is easier
  to visualize but involves several levels of range searching structures. The running
  time of this approach, which we do not detail here, is asymptotically the same as 
  the bound that we get here.}

Fix an arc $\gamma \in \Gamma$, and recall that it is a constant-degree algebraic arc in the plane,
and, by assumption, $\gamma$ is convex, smooth and $x$-monotone. 
Consider the smallest affine variety (curve) $V_\gamma$ that contains $\gamma$, 
known as the \emph{Zariski closure} of $\gamma$~\cite{CLO}, and let $F(x,y)$ 
be the bivariate polynomial whose zero set is $V$. 
$F$ is a polynomial of constant degree, which we denote by $d$.
Consider a line $\ell$, given by the equation $y := ax + b$, where $a, b$ are real coefficients.
(Vertical lines are easier to handle, and we ignore this special case in what follows.)
Then $\ell$ intersects $\gamma$ if and only if there exists $x\in\reals$ such that $(x,ax+b)\in\gamma$.
This can be expressed as a quantified Boolean algebraic predicate of constant complexity
(i.e., involving a constant number of variables, and a constant number of polynomial
equalities and inequalities of constant degrees); one of the clauses of the predicate is
$F(x,ax+b) = 0$ and the others restrict $(x,ax+b)$ to lie in $\gamma$.
%
%
Using the singly exponential quantifier-elimination algorithm in~\cite[Theorem 14.16]{BPR} 
(also used earlier in Section~\ref{sec:wide}), we can construct, in $O_d(1)$ time, a quantifier-free 
semi-algebraic set $G := G_{\gamma}$ in the $ab$-parametric plane, whose overall complexity 
is $O_d(1)$ as well, such that the quantified predicate is true
if and only if $(a,b) \in G$; see, e.g.,~\cite{AM-94} for a concrete construction of such 
a set for the problem of intersection detection between lines and spheres in $\reals^3$.

We have thus mapped the setting of our problem to a planar point location problem amid
a collection $\G$ of $n$ semi-algebraic regions of constant complexity. Using standard 
techniques based on $\eps$-cuttings 
(see, e.g.,~\cite{CF-90} for such constructions), 
one can construct, using overall storage and preprocessing time of $O(n^{3/2+\eps})$,
for any $\eps > 0$, a data structure that supports point-location queries in the arrangement
$\A(\G)$ of these regions in time $O(n^{1/2})$ per query;\footnote{%
  We comment that we need to exploit our model of computation in which root extraction,
  and manipulations of these roots, can be performed in constant time, for 
  univariate polynomials of constant degree.}
note that the factor $n^\eps$ does not appear in the query time bound, but only in the preprocessing time bound. 

Briefly, to do so, we construct a $(1/\sqrt{n})$-cutting of $\A(\G)$ in $O(n^{3/2})$ time.
This is a partition of the plane into $O(n)$ pseudo-trapezoids, each crossed by $O(n^{1/2})$
boundaries of the regions in $\G$ (see \cite{CF-90}). Each pseudo-trapezoid (trapezoid for short)
has a conflict list $\G_tau$ of the set of regions whose boundaries cross $\tau$, and another
list $\G^{(0)}_\tau$ of regions that fully contain $\tau$. The lists $\G_\tau$ are stored
explcitly at the respective regions $\tau$, as their overall size is $O(n^{3/2})$.
The overall size of the lists $\G^{(0)}_\tau$ is $O(n^2)$, so we store them implicitly
in a persistent data structure, based on some tour of the trapezoids of the cutting, 
using the fact that $\G^{(0)}_\tau$ changes by only $O(n^{1/2})$ regions as we pass
from $\tau$ to an adjacent trapezoid $\tau'$ (we gloss here over certain technical issues
involved in the construction of such a tour).
explcitly at the respective regions $\tau$, as their overall size is $O(n^{3/2})$.

For the problem of detecting line intersection, it suffices to test whether 
the query point $(a,b)$ (representing a query line $\ell$) is contained in any 
of the input semi-algebraic regions. To do so, we locate $(a,b)$ in the cutting.
If its containing trapezoid $\tau$ has a nonempty list $\G^{(0)}_\tau$, we
report that $\ell$ intersects an arc of $\Gamma$ and stop. Otherwise we go over
the conflict list $\G_\tau$ and test explicitly whether $\ell$ interscts any
of the associated arcs of $\Gamma$, in $O(n^{1/2})$ time.
Moreover, this point-location machinery can also return a compact representation 
of the set of the arcs from $\Gamma$
that intersect $\ell$, as a disjoint union of $O(n^{1/2})$ precomputed 
canonical subsets of $\Gamma$ (namely, the set $\G^{(0)}_\tau$ of the trapezoid
$\tau$ containing $(a,b)$, and the $O(n^{1/2})$ singleton sets corresponding
to those regions in $\G_\tau$ that contain $(a,b)$). This latter property is 
useful for the extensions of this procedure for detecting intersections of 
rays or segments with the given arcs, described below.

\subsubsection{Detecting ray and segment intersections with the arcs}
\label{sec:ray}

We next augment the data structure so that it can test whether
a query ray $\rho$ intersects any arc in $\Gamma$. 
A similar, somewhat more involved approach,
which is spelled out later in this section, allows us
also to test whether a query segment $s$ intersects any arc in $\Gamma$. 
Using the parametric-search machinery of Agarwal and Matou\v{s}ek~\cite{AM}, 
this latter structure allows us to answer ray shooting queries (finding the 
first arc of $\Gamma$ hit by a query ray $\rho$) with similar performance bounds.

We comment that in principle we could have simply used an extended version
of the quantifier-elimination technique used in the previous subsection.
However, such an extension requires more parameters to represent a ray
(three parameters) or a segment (four parameters). As a consequence, the
space in which we need to perform the search becomes three- or four-dimensional,
and the performance of the solution deteriorates. We therefore use a different,
more explicit approach to these extended versions.

We also comment that the analysis presented next only applies to nonvertical 
rays and segments. Handling vertical rays is much simpler, and amounts, with 
some careful modifications, to point location of the apex of the ray in the 
arrangement of the given arcs, which can be implemented with standard techniques, 
with performance bounds that match the ones that we obtain for the general problem.
We therefore assume in what follows that the query rays and segments are nonvertical.

\begin{figure}[htb]
  \begin{center}
    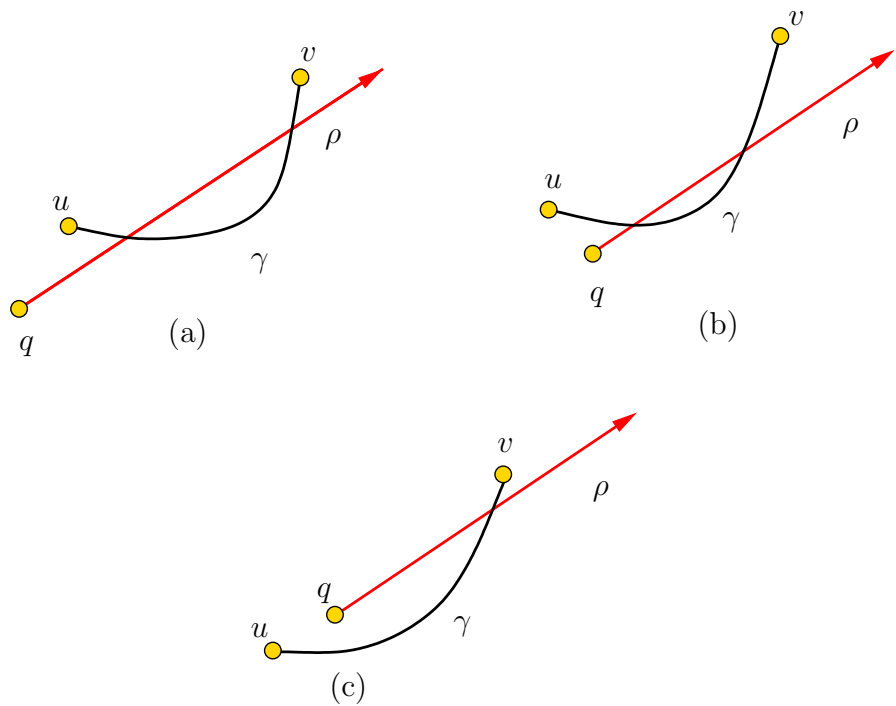
    \caption{\sf{Scenarios where a ray $\rho$ intersects a convex $x$-monotone arc $\gamma$: 
        (a) $q$ lies to the left of $u$.
        (b) $q$ lies between $u$ and $v$ and below $\gamma$, and
        the tangent direction to $\gamma$ at $q_x$ is smaller than $a$.
        (c) $q$ lies between $u$ and $v$ and above $\gamma$, and $v$ lies above $\ell$.}}
    \label{rho-cross}
  \end{center}
\end{figure}

So let $\rho$ be a query ray, let $q$ be the apex of $\rho$,
and let $\ell$ be the line supporting $\rho$. We assume, without loss of 
generality, that $\rho$ is directed to the right (for rays directed to 
the left, a symmetric set of conditions apply). We have:

\begin{lemma}
  \label{lem:ray-arc}
  Let $\rho$, $q=(q_x,q_y)$ and $\ell$ be as above. Then $\rho$ intersects 
  a convex $x$-monotone arc $\gamma$, oriented from left to right, 
  if and only if $\ell$ intersects $\gamma$, and one of the following
  conditions holds, where $u$ and $v$ are the left and right endpoints 
  of $\gamma$, and where $a$ is the slope of $\ell$.
  \begin{description}
  \item[(a)]
    $q$ lies to the left of $u$.
    See Figure~\ref{rho-cross}(a).
  \item[(b)]
    $q$ lies between $u$ and $v$ and below $\gamma$, and
    the tangent direction to $\gamma$ at $q_x$ is smaller than $a$.
    See Figure~\ref{rho-cross}(b).
  \item[(c)]
    $q$ lies between $u$ and $v$ and above $\gamma$, and $v$ lies above $\ell$.
    See Figure~\ref{rho-cross}(c).
  \end{description}
\end{lemma} 

\noindent{\bf Proof.}
The `only if' part of the lemma is simple, and we only consider the `if' part.
We are given that $\ell$ intersects $\gamma$. If $q$ lies to the left of $u$ then
clearly $\rho$ also intersects $\gamma$ (this is Case (a), where we actually have 
$\ell \cap \gamma = \rho \cap \gamma$), and if $q$ lies to the 
right of $v$ then clearly $\rho$ does not intersect $\gamma$. 
Assume then that $q$ lies between $u$ and $v$. If $q$ lies above $\gamma$, 
the ray intersects $\gamma$ if and only if $v$ lies above $\ell$, as is easily checked,
which is Case (c). If $q$ lies below $\gamma$ then, given that $\ell$ intersects $\gamma$, 
$\rho$ intersects $\gamma$ if and only if $q$ lies to the left of the left intersection
point in $\ell\cap\gamma$, and this happens if and only if the slope of the tangent
to $\gamma$ at $q_x$ is smaller than the slope of $\ell$. This is Case (b), and thus
the proof is completed.
$\Box$

\medskip
Our data structure is constructed by taking the structure of Section~\ref{sec:line}
and augmenting it with additional levels, in three different ways, each testing for 
one of the conditions (a), (b), (c) in Lemma~\ref{lem:ray-arc}. 

Testing for Condition (a) is easily done with a single additional level 
based on a one-dimensional search tree on the left endpoints of the arcs.

Testing for Condition (b) requires three more levels.
The first level is a segment tree on the $x$-spans of the arcs, which we 
search with $q_x$, to retrieve all the arcs whose $x$-span contains $q_x$, as
the disjoint union of $O(\log n)$ canonical sets.
The second level filters out those arcs that lie above $q$.
As in the line-intersection structure given in Section~\ref{sec:line}
(except that the plane in which we perform the search is the actual
$xy$-plane and not the parametric $ab$-plane),
this level requires $O(n^{3/2+\eps})$ storage and preprocessing (where 
$n$ is the size of the present canonical set),
and answers a query in $O(n^{1/2})$ time.
In the third level we consider the tangent directions of the arcs of $\gamma$
as partial functions of $x$, and construct their lower envelope (see~\cite[Corollary 6.2]{SA}).
We can then test whether $(q_x,a)$ lies above the envelope in logarithmic time.

Finally, testing for Condition (c) also requires three more levels, where the first 
two levels are as in case (b), and the third level tests whether there is any right
endpoint $v$ of an arc in the present canonical set that lies above $\ell$, by constructing, 
in nearly linear time, the upper convex hull of the right endpoints and by testing, in 
logarithmic time, whether $\ell$ does not pass fully above the hull (see, e.g.,~\cite{DK}).

It is easily verified that the overall data structure has the desired performance
bounds, namely, $O(n^{3/2+\eps})$ storage and preprocessing cost, and $O(n^{1/2+\eps})$ 
query time, for any $\eps > 0$.

\paragraph{Detecting segment intersection.}
The same mechanism works when $\rho$ is a segment, rather than a ray, except 
that the conditions for intersection with an arc of $\Gamma$ are more involved.
To simplify the presentation, we reduce the problem to the ray-intersection detection
problem just treated, thereby avoiding explicit enumeration of quite a few subcases that
need to be tested.

\begin{figure}[htb]
  \begin{center}
    \begin{tabular}{c c}
    {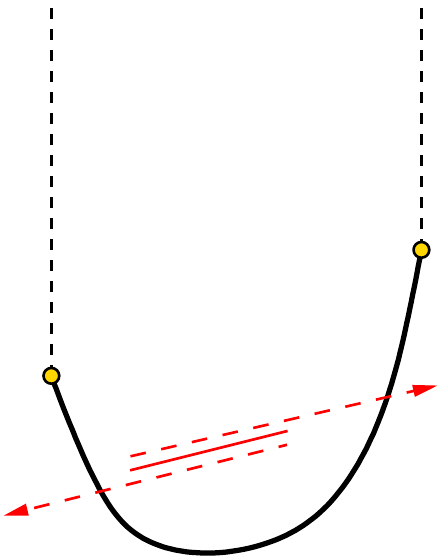 }  &\quad
    {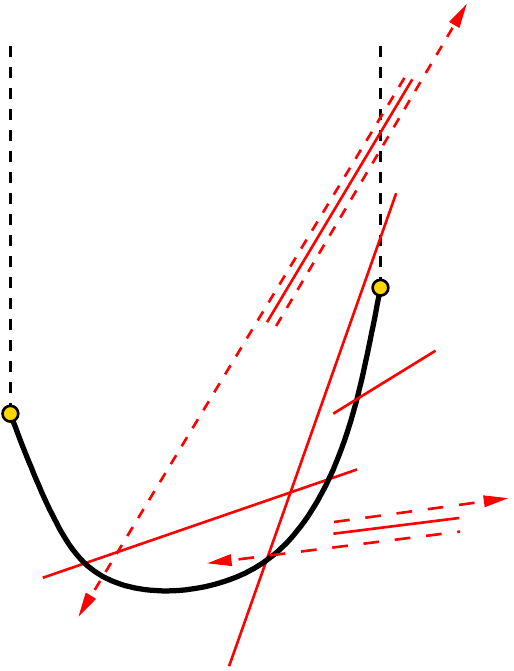 }  \\
    {\small (a)} & \quad\quad
    {\small (b)} 
    \end{tabular}
  \end{center}
  \caption{\sf{A segment $s$ and a convex $x$-monotone arc $\gamma$, where the line $\ell$ containing $s$ intersects
      $\gamma$: 
      (a) Both endpoints of $s$ lie inside $\kappa$ and therefore $s$ is disjoint from $\gamma$,
      although in this case $\rho_p$, $\rho_q$ (depicted by the dash arrows in the figure) both intersect $\gamma$.
      (b) Several different positions of $s$, in each of which there is an endpoint of $s$ outside $\kappa$.
      In this case $s$ intersects $\gamma$ if and only if both $\rho_p$ and $\rho_q$ intersect $\gamma$ 
      (these rays are drawn for an illustration for only two of the segments in the subfigure).
    }
  }
  \label{seg-cross}
\end{figure}

We associate with each arc $\gamma\in\Gamma$ the semi-unbounded region 
\[
\kappa = \kappa(\gamma) = \{ (x,y) \mid u_x \le x \le v_x \text{ and $(x,y)$ lies strictly above $\gamma$} \} .
\]
That is, $\kappa$ is bounded on the left by the
upward vertical ray emanating from $u$, bounded on the right by the
upward vertical ray emanating from $v$, and bounded from below by $\gamma$; 
see Figure~\ref{seg-cross}.
Then we have the following extension of Lemma~\ref{lem:ray-arc}:
\begin{lemma}
  \label{lem:seg-arc}
  Let $\gamma$, $u$, $v$, and $\kappa$ be as above.
  Let $s$ be a segment, with left endpoint $p=(p_x,p_y)$, right endpoint
  $q=(q_x,q_y)$, and slope $a$, and let $\ell$ be the line supporting $s$. 
  Let $\rho_p$ be the ray that starts at $p$ and contains $s$ (so $\rho_p$ 
  is rightward directed), and let $\rho_q$ be the ray that starts at $q$ 
  and contains $s$ (so $\rho_q$ is leftward directed).
  Then $s$ intersects $\gamma$ if and only if all the following conditions hold:
  \begin{description}
  \item[(a)] $\ell$ intersects $\gamma$.

  \item[(b)] At least one of $p$, $q$ lies outside $\kappa$.

  \item[(c)] $\rho_p \cap \gamma \neq \emptyset$ and $\rho_q \cap \gamma \neq \emptyset$.
  \end{description}
\end{lemma} 
\medskip
See Figure~\ref{seg-cross} for an illustration.

\medskip

\noindent{\bf Proof.}
Here too, the `only if' part of the lemma is simple, and we only consider the `if' part.
Condition (a) and the convexity of $\gamma$ imply that $\ell\cap \gamma$ consists of one or two
points. 

Assume first that $\ell\cap \gamma$ consists of two points $\xi$, $\eta$. By Condition (b)
at least one of $p$, $q$ lies outside the (open) interval $\xi\eta$. Assume, without loss of generality,
that $p$ lies outside $\xi\eta$. If $p$ lies to the right of that interval, $\rho_p$ misses $\gamma$,
contradicting Condition (c). Thus $p$ lies to the left of $\xi\eta$. Then the only way in which
$s\cap\gamma$ is empty is when $q$ also lies to the left of $\xi\eta$, but then $\rho_q\cap\gamma$
would be empty, again contradicting Condition (c). 

Assume next that $\ell\cap \gamma$ consists of one point $\xi$. By Condition (c), 
$p$ must lie to the left of $\xi$ and $q$ must lie to the right of $\xi$, implying that $s$
meets $\gamma$. See Figure~\ref{seg-cross}(b) for an illustration.
$\Box$

The description of the data structure is fairly straightforward given the 
criteria for intersection in Lemma~\ref{lem:seg-arc}.
That is, we construct a multi-level data structure, where in the first level 
we test Condition (a), obtaining the set of arcs that $\ell$ intersects, as the disjoint union of
canonical sets of arcs. At the next levels we test Condition (b). For an arc $\gamma$, with
endpoints $u$ and $v$, a point $p$ lies outside $\kappa = \kappa(\gamma)$ when $p$ lies either
below $\gamma$, or to the left of $u$, or to the right of $v$. Collecting the arcs $\gamma$
that satisfy this property is easily done using similar data structures as those described 
for the case of ray-intersection queries, where we first extract those arcs that lie above $p$ and
then the arcs that lie above $q$. We then use a one-dimensional search tree on the left (resp., right) 
endpoints of the arcs, to collect those arcs that lie to the right of $p$ (resp., to the left of $q$).
Overall this requires $O(n^{3/2+\eps})$ storage and preprocessing and the query time is $O(n^{1/2+\eps})$.
To test for Condition (c), we build a multi-level data structure, over each final canonical set,
that tests whether Conditions (a)--(c) of Lemma~\ref{lem:ray-arc} are satisfied for the
rightward-directed ray $\rho_p$, and that their symmetric counterparts are satisfied for
the leftward-directed ray $\rho_q$. 

The overall performance bounds remain the same: $O(n^{3/2+\eps})$ storage and preprocessing cost,
and $O(n^{1/2+\eps})$ query time.


As noted above, the parametric-search approach in~\cite{AM} yields a procedure
for answering ray-shooting queries, given a procedure for answering segment-intersection queries,
as well as a parallel procedure for the same task.
The preprocessing that constructs the structure is performed only sequentially, as above.
The query procedure, for detecting segment intersection, is easy to parallelize, since 
it is essentially a multi-level tree traversal. By allocating a processor to each node that
the query visits, we can perform the traversal in parallel, in polylogarithmic parallel time.
In other words, the parallel time to answer a 
segment-intersection detection query is $O(\polylog{n})$, using at most 
$O(n^{3/2+\eps})$ processors.
Integrating these bounds with~\cite[Theorem 2.1]{AM}, we obtain that ray-shooting queries 
in a planar collection of arcs can be answered using the
same data structure for segment-intersection queries, where the query time 
for the former problem is only within a polylogarithmic factor of the 
time for the latter one, a factor that is hidden in our $\eps$-notation, by slightly increasing $\eps$.

A simple modification of the segment-intersection query procedure allows
us to report an arc $\gamma$ intersecting (i.e., containing) the endpoint of 
the query segment, when the segment is otherwise empty. The easy details are omitted.

In conclusion we have shown the following general result, which we believe
to be of independent interest.
\begin{proposition}
  \label{prop:ray-curves}
  A collection $\Gamma$ of $n$ constant-degree algebraic arcs in the plane can be
  preprocessed, in time and storage $O(n^{3/2+\eps})$, for any $\eps>0$, into a
  data structure that supports ray shooting queries in $\Gamma$, in
  $O(n^{1/2+\eps})$ time per query.
\end{proposition}
As a corollary, we thus obtain:
\begin{proposition}
  \label{prop:onzf}
  For a partitioning polynomial $f$ of sufficiently large constant degree, and a 
  set $\W$ of $n$ triangles, one can construct a data structure of size and 
  preprocessing cost $O(n^{3/2+\eps})$, so that a segment-shooting query with a
  segment that lies in $Z(f)$, can be answered in $O(n^{1/2+\eps})$ time, for any $\eps > 0$.
\end{proposition}
As already concluded, Proposition~\ref{prop:onzf}, combined with Proposition~\ref{prop:inside}
of the previous subsection, complete the proof of Theorem~\ref{thm:trimain}.

\section{Segment-triangle intersection reporting, emptiness, and approximate counting queries}
\label{sec:seg}

\subsection{Segment-triangle intersection reporting and emptiness}

We extend the technique presented in Section~\ref{sec:shoot}
to answer intersection reporting queries amid triangles in $\reals^3$. 
Here too we have a set $\T$ of $n$ triangles in $\reals^3$, and our goal is to
preprocess $\T$ into a data structure that supports efficient intersection
queries, each of which specifies a line, ray or segment $\rho$ and asks for 
reporting the triangles of $\T$ that $\rho$ intersects. In particular, this
data structure also supports \emph{segment emptiness} queries, in which we 
want to determine whether the query segment meets any input triangle. 
Unfortunately, for technical reasons, the method does not extend to
segment-triangle intersection (exact) \emph{counting} queries, in which we want to
find the (exact) number of triangles that intersect a query segment (or a line or a ray).
This issue will be discussed later on in this section, and a partial solution,
which supports queries that approximately count the number of intersections, up to
any prescribed relative error $\eps>0$, will be presented in Section~\ref{sec:apxct}.

\begin{theorem}
  \label{thm:seg_intersection}
  Given a collection of $n$ triangles in three dimensions, and
  a prescribed parameter $\eps>0$, we can process the triangles into a data structure
  of size $O(n^{3/2+\eps})$, in time $O(n^{3/2+\eps})$, so that a segment-intersection 
  reporting (resp., emptiness) query amid these triangles can be answered in 
  $O(n^{1/2+\eps}+k\log n)$ (resp., $O(n^{1/2+\eps})$) time, 
  where $k$ is the number of triangles that the query segment crosses.
\end{theorem}

\noindent{\bf Proof.}
The algorithm that we develop here is a fairly easy adaptation of those given 
in the previous section, which is fairly straightforward, except
for one significant issue, noted below. The preprocessing is almost identical, 
except that now we preprocess the sets $\T_C$ of wide triangles for line- 
(ray-, or segment-)intersection reporting queries in a set of planes in 
$\reals^3$ (namely, the corresponding planes in $\H_C$); in this case the reporting 
query time is ${\displaystyle O\left(\frac{n \polylog{n} }{s^{1/3}} + k\right)}$,
where $n$ is the number of wide triangles, $s$ is the amount of storage 
allocated, and $k$ is the output size; see~\cite[Theorem 3.3]{AM}.

To answer a query with a segment (ray, or line) $\rho$, we
trace $\rho$ through the cells and subcells that it crosses, as before,
but do not abort the search at cells where an intersection has been found,
and instead follow the search to completion. We take each canonical set
$\T_C$ that the query collects, and access it via the intersection-reporting
mechanism that we have constructed for it. At the bottom level we examine
all the triangles in the subproblem, and report those that are crossed by $\rho$.

Recall that the canonical sets that we construct are not necessarily pairwise 
disjoint, even those that are encountered when querying with a fixed segment $e$. 
Thus, the triangles that we report may be reported multiple times, which we want 
to avoid. To this end, at each level of the recursion (on the trapezoidal cells $\psi$)
within a single cell $\tau$ of the polynomial partition, we take the outputs of the two 
recursive calls (recall the details in Section~\ref{sec:shoot}),
and keep only one copy of each triangle that is reported twice. (Recall that
the sets of triangles in the recursive calls are disjoint from the canonical 
set $\T_C$ but may share triangles between themselves.)
Repeating this step at all levels of recursion guarantees that the reported
triangles are all distinct. The overall cost of this overhead is $O(k\log n)$,
where $k$ is the output size.

Note that this non-disjointness of the outputs makes it difficult to convert
this procedure to one that counts the number of triangles that the query object
crosses, and at the moment we do not know how to perform intersection-counting
queries with the same performance bounds. (It is also conceivable that the
upper bound on the cost of a counting query is larger; see, e.g., \cite{AM} 
for similar phenomena.)

A similar adaptation is applied for the substructure that handles queries
that are contained in the zero set of the partitioning polynomial, and
its easy details are omitted.
(In fact, in this case, due to the nature of our range searching mechanism, 
the conflict lists comprising the answer to a single query are pairwise disjoint. 
Therefore in this case, we do obtain a range counting mechanism with similar 
asymptotic performance bounds).

This completes the description of the required adaptations, and 
establishes Theorem~\ref{thm:seg_intersection}.
$\Box$

\subsection{Approximate segment-intersection counting} \label{sec:apxct}

Let $T$ be as above, and let $\delta>0$ be a prescribed parameter.
We want to preprocess $T$ into a data structure that supports
\emph{approximate counting} queries of the following form: Given a 
query segment $e$, count how many triangles of $T$ are crossed by $e$, 
up to a relative error of $1\pm \delta$.

To do so, we use the notion of a \emph{relative $(p,\delta)$-approximation},
as developed and analyzed in Har-Peled and Sharir~\cite{HPS}.
We recall this notion and its basic properties.
Consider the range space $(T,\R)$, where $\R$ is the collection of 
all subsets of $T$ of the form $T(e) = \{\Delta\in T \mid \Delta\cap e\ne\emptyset\}$,
where $e$ is a segment. It is easily shown that $(T,\R)$ has finite VC-dimension $\eta$.

(A brief argument for this latter property follows by bounding the \emph{primal shatter function} 
of the range space, as a function of $|T|$. This is done by representing the lines supporting 
the triangle edges as surfaces in $4$-space, and by observing that, for each of the
$O(|T|^4)$ cells $C$ of their arrangement, all the lines whose Pl\"ucker images
lie in $C$ meet the same subset $T_C$ of triangles of $T$. For a segment $e$,
we take the cell $C$ containing the image of the line supporting $e$, and argue
that there are only polynomially many subsets of $T_C$ that can be crossed by such 
a segment $e$.)

For a segment $e$ and a subset $X\subseteq T$, write $\oX(e) := \frac{|X\cap e|}{|X|}$;
this is the ``relative size'' of the range $T(e)$ induced by $e$. Let $0<p,\;\delta < 1$ be given parameters.
A subset $Z\subseteq T$ is called a \emph{relative $(p,\delta)$-approximation} if, for every segment $e$, we have:
\begin{gather*}
  (1-\delta)\oT(e) \le \oZ(e) \le (1+\delta)\oT(e), \quad\text{if $\oT(e) \ge p$} \\
  \oT(e) - \delta p \le \oZ(e) \le \oT(e) + \delta p, \quad\text{if $\oT(e) \le p$} .
\end{gather*}
As shown in \cite{HPS}, a random sample $Z$ of $T$ of size 
${\displaystyle \frac{c}{\delta^2 p} \left( \eta \log \frac{1}{p} + \log \frac{1}{q} \right)}$
is a relative $(p,\delta)$-approximation with probability at least $1-q$, 
where $c$ is some absolute constant.

We use this notion as follows. Using Theorem~\ref{thm:seg_intersection}
we construct our data structure for exact segment intersection \emph{reporting} queries,
with $O(n^{3/2+\eps})$ storage and preprocessing and query time $O(n^{1/2+\eps} + k\log n)$,
for any $\eps>0$, where $k$ is the output size. We also construct
a relative $(p,\delta)$-approximation $Z$ for $T$, by an appropriate
random sampling mechanism, where the value of $p$ will be determined shortly. 

An approximate counting query with a segment $e$ is answered as follows. 
We first query with $e$ in the data structure for exact segment intersection 
reporting, but stop the procedure as soon as we collect more 
than $np$ triangles. If the output size $k$ does not exceed this 
bound, we output $k$ (as an exact count) and are done. 
The cost of the query so far is $O(n^{1/2+\eps} + np\log n)$.

If we detect that $k > np$, we compute $\oZ(e)$ by brute force,
in $O(|Z|)$ time, and output the value $k_{\rm apx} := n\oZ(e)$.
By the properties of relative approximations, we have, since
$\oT(e) \ge p$,
\[
(1-\delta) k \le k_{\rm apx} \le (1+\delta) k ,
\]
so $k_{\rm apx}$ satisfies the desired approximation property.

The overall (deterministic) cost of the query is
\[
O\left(n^{1/2+\eps} + np\log n + |Z|\right) =
O\left(n^{1/2+\eps} + np\log n + 
\frac{1}{\delta^2 p} \left( \eta \log \frac{1}{p} + \log \frac{1}{q} \right) \right) .
\]
We ignore the effect of $q$, and balance the terms by choosing
${\displaystyle p := \frac{1}{\delta n^{1/2}}}$, making the query cost
\[
O\left(n^{1/2+\eps} + \frac{n^{1/2}}{\delta} \log n \right) .
\]
As long as $\delta$ is not too small (but we can still choose $\delta$ 
to be $1/n^{\eps'}$, for some $\eps' < \eps$), the first term dominates the bound,
which is thus asymptotically the same as the bound for reporting queries.

The storage is $O\left( n^{3/2+\eps} + |Z|\right) = O\left( n^{3/2+\eps}\right)$,
as long as $\delta$ is not chosen too small.

We thus conclude:

\begin{theorem}
  \label{thm:approximate_counting}
  Given a collection of $n$ triangles in three dimensions, and
  prescribed parameters $\eps, \delta>0$, where $\delta = \omega(1/n^\eps)$,
  we can process the triangles, using random sampling, into a data structure
  of size $O(n^{3/2+\eps})$, in time $O(n^{3/2+\eps})$, so that, for a query segment
  $e$, the number of intersections of $e$ with the input triangles can be 
  approximately computed, up to a relative error of $1\pm\delta$, with very
  high probability, in $O(n^{1/2+\eps})$ time.
\end{theorem}

\section{Tradeoff between storage and query time}
\label{sec:trade}

In this section we extend the technique in Sections~\ref{sec:shoot} and \ref{sec:seg} to
obtain a tradeoff between storage (and preprocessing) and query time.
A similar tradeoff holds for the other problems studied in Section~\ref{sec:other}.

For a quick overview of our approach, consider the ray-shooting structure of 
Section~\ref{sec:shoot}, and let $s$ be the storage parameter that we allocate 
to the structure, which now satisfies $n\le s\le n^4$. We modify the procedure 
for ray shooting inside a cell $\tau$ by (i) stopping potentially the $r_0$-recursion 
at some earlier `premature' level, and (ii) modifying the structure at the bottom 
of recursion so that it uses the (weaker) ray-shooting technique of 
Pellegrini~\cite{Pel} instead of a brute-force scanning of the triangles 
(the current cost of $O(n^2/s)$, a consequence of this brute-force approach,
is too expensive when $s$ is small). A similar
adaptation is applied to the procedure of ray shooting on the zero set of the
partitioning polynomial. With additional care 
we obtain the performance bounds (\ref{eq:trade1}) and 
(\ref{eq:trade2}) announced in the introduction.

We now present the technique in detail.
Consider the ray-shooting structure of Section~\ref{sec:shoot}, and let 
$s$ be the storage parameter that we allocate to the structure, which
satisfies $n\le s\le n^4$. As before, we use this notation to indicate 
that the actual storage (and preprocessing) that the structure uses may be $O(s^{1+\eps})$, 
for any $\eps>0$. We comment that in the preceding sections $s$ is assumed 
to be at most $n^2$. Handling larger values of $s$ require some care, 
detailed below. For the time being, we continue to assume that $s\le n^2$, 
and will later show how to extend the analysis for larger values.

Consider first the subprocedure for handling ray shooting for rays that
are not contained in the zero set of the partitioning polynomial.
%
When $s = O(n^{3/2})$, we run the recursive preprocessing described in 
Section~\ref{sec:shoot} up to some `premature' level $k$, and when 
$s = \Omega(n^{3/2})$, we run it all the way down. 
With a suitable choice of parameters, 
we obtain $O(D^{3k+\eps})$ subproblems at the bottom level of recursion, 
each involving at most $n/D^{2k}$ (narrow) triangles. 

Except for the bottom level, 
we build, at each node $\tau$ of the recursion, the same structure on 
the set $\W_\tau$ of wide triangles in $\tau$, with one (significant) 
difference. First, since we start the recursion with storage parameter 
$s$, we allocate to each subproblem, at any level $j$, the storage 
parameter $s/D^{3j}$, thus ensuring that the storage used by the 
structure is $O(s^{1+\eps})$. However, the cost of a query, even at the first 
level of recursion, given in (\ref{eq:qns}), has the term $O(n^{2+\eps}/s)$,
which is the cost of a naive, brute-force processing of the conflict lists 
at the bottom instances of the $r_0$-recursion within the partition cells.
This is fine for $s = \Omega(n^{3/2+\eps})$ but kills the efficiency of the
procedure when $s$ is smaller. For example, for $s=n$ we get (near) linear 
query time, much more than what we aim to have. We therefore improve
the performance at the bottom-level nodes of the $r_0$-recurrence (within a cell), 
by constructing, for each respective conflict list, the ray-shooting data 
structure of Pellegrini~\cite{Pel}, which, for $N$ triangles and with 
storage parameter $s$, answers a query in time $O(N^{1+\eps}/s^{1/4})$.
Since at the bottom of the $r_0$-recursion, both the number of triangles
and the storage parameter are $n^{2+\eps}/s$,
the cost of a query at the bottom of the recursion is $O((n^2/s)^{3/4+\eps})$.
That is, the modified (improved) cost of a query at such a node is
\begin{equation}
  \label{eq:qns1}
  Q(n,s) = O\left( \frac{n \ {\rm polylog}(n) }{s^{1/3}} + \frac{n^{3/2+\eps}}{s^{3/4}} \right) .
\end{equation}

At each of the $O(D^{3k+\eps})$ bottom-level cells $\tau$, we take the 
set $\N_\tau$ of (narrow) triangles that have reached $\tau$, whose size is now 
at most $n/D^{2k}$, allocate to it the storage parameter $s/D^{3k}$, 
and preprocess $\N_\tau$ using the aforementioned technique of 
Pellegrini~\cite{Pel}, which results in a data structure, with storage 
parameter $s/D^{3k}$, which supports ray shooting queries in time 
\[
O\left(\frac{|\N_\tau|^{1+\eps}}{(s/D^{3k})^{1/4}}\right) = 
O\left(\frac{(n/D^{2k})^{1+\eps}}{(s/D^{3k})^{1/4}}\right) = 
O\left(\frac{n^{1+\eps}}{s^{1/4}D^{(5/4+2\eps)k}} \right) .
\]
Multiplying this bound by the number $O(D^{k+\eps})$ of cells that the
query ray crosses, the cost of the query at the bottom-level cells is 
\[
Q_{\rm bot}(n,s) = O\left(\frac{n^{1+\eps}}{s^{1/4}D^{(1/4+\eps)k}} \right) .
\]
The cost of a query at the inner recursive nodes of some depth $j<k$
is the number, $O(D^{j+\eps})$, of $j$-level cells that the ray crosses, 
times the cost of accessing the data structure for the wide triangles 
at each visited cell. Since we have allocated to each of the 
$O(D^{3j+\eps})$ cells at level $j$ the storage parameter $O(s/D^{3j})$,
the cost of accessing the structure for wide triangles in a $j$-level
cell is, according to (\ref{eq:qns1}), at most
\[
Q_{\rm inner}(n,s) = O\left( \frac{(n/D^{2j}){\rm polylog}(n)}{(s/D^{3j})^{1/3}} + 
\frac{(n/D^{2j})^{3/2+\eps}}{(s/D^{3j})^{3/4}} \right) = 
O\left( \frac{n\;{\rm polylog}(n)}{ D^{j} s^{1/3} } +
\frac{n^{3/2+\eps}}{D^{(3/4+2\eps)j}s^{3/4}} \right) . 
\]
Summing over all $j$-level cells, for all $j$, and then 
adding the bottom-level cost, and the cost of traversing 
the structure with the query segment, the overall cost of a query is
(we remind the reader that so far we only consider the case where $s\le n^2$)
\begin{equation}
  \label{qtrade}
  O\left( D^{k+\eps} + \frac{n^{3/2+\eps} D^{k/4} }{s^{3/4}} + \frac{n^{1+\eps}}{s^{1/3}} + \frac{n^{1+\eps}}{s^{1/4}D^{(1/4+\eps)k}} \right) .
\end{equation}
We choose $k$ to (roughly) balance the second and the last terms; specifically, we choose
\[
D^k = \frac{s}{n} .
\]
Since $D^{k+\eps}$ should not exceed $O(n^{1/2+\eps})$, we require for this choice of $k$
that $s = O(n^{3/2+\eps})$. In this case it is easily verified that the second and last terms, 
which are $O(n^{5/4+\eps}/s^{1/2})$, dominate both the first and third terms (recall that we 
assume $s \ge n$), and the query time is therefore
\[
O(n^{5/4+\eps}/s^{1/2}) .
\]
For larger values of $s$, that is, when $s = \Omega(n^{3/2+\eps})$ (but we still
assume $s\le n^2$), we balance the first term with the last term, so we choose
\[
D^k = \frac{n^{4/5}}{s^{1/5}} .
\]
Note that in this range we indeed have that $D^{k+\eps} = O(n^{1/2+\eps})$.
Moreover, in this case the first and last terms dominate the second and 
third terms, as is easily verified. Therefore the query time is
\[
O(n^{4/5+\eps}/s^{1/5}) .
\]
As already promised, the case where the query ray lies on the current zero set will be presented later.

It remains to handle the range $n^2 < s\le n^4$.
Informally, at each cell $\tau$ of the polynomial partition, at any level
$j$ of the $D$-recursion, we have $n_\tau\le n/D^{2j}$ wide triangles
and storage parameter $s_\tau = s/D^{3j}$. Since $s\ge n^2$, we also have
$s_\tau \ge n_\tau^2$. With such `abundance' of storage, we run the 
$r_0$-recursion until we reach subproblems of constant size, in which 
case we simply store the list of wide triangles at each bottom-level node,
and the query simply inspects all of them, at a constant cost per subproblem. Hence
the cost of a query at $\tau$ is $O(n_\tau^{1+\eps}/s_\tau^{1/3})$.
To be precise, this is the case as long as $s_\tau \le n_\tau^3$.
If $n^2\le s\le n^3$ there will be some level $j$ at whose cells $\tau$
$s_\tau = s/D^{3j}$ becomes larger than $(n/D^{2j})^3 \ge n_\tau^3$, and then 
the cost becomes $O(n_\tau^\eps)$.
When $n^3 < s \le n^4$ the cost becomes $O(n_\tau^\eps)$ right away (and stays so). That is, the cost 
of a query in the structure for wide triangles at a cell $\tau$ at level $j$ is
\begin{align*}
O\left( \frac{(n/D^{2j})^{1+\eps}} {(s/D^{3j})^{1/3}} \right) = 
O\left( \frac{n^{1+\eps}} {s^{1/3} D^{j(1+2\eps)}} \right) & \qquad\text{for $s \le \frac{n^3}{D^{3j}}$} \\ 
O\left( n^\eps \right) & \qquad\text{for $s > \frac{n^3}{D^{3j}}$} .
\end{align*}
Since a query visits $O(D^{j+\eps})$ cells $\tau$ at level $j$, 
the overall cost of searching amid the wide triangles, over all levels, 
is easily seen to be
\begin{align*}
O\left( \frac{n^{1+\eps}} {s^{1/3}} \right) & \qquad\text{for $n^2\le s \le n^3$} \\ 
O\left( D^k n^\eps \right) & \qquad\text{for $n^3< s \le n^4$} ,
\end{align*}
where $k$ is the depth of the $D$-recursion.

Querying amid the narrow triangles is done again as in Section~\ref{sec:shoot} (once again, recall that we now
consider the case where $s > n^2$, whereas earlier in this section we assumed $s \le n^2$).
At each node $\tau$ at the bottom level $k$ of the $D$-recursion we use 
Pellegrini's data structure~\cite{Pel}, which, with at most $n/D^{2k}$
narrow triangles and storage parameter $s/D^{3k}$, answers a query in time 
\[
O\left( \frac{(n/D^{2k})^{1+\eps}} {(s/D^{3k})^{1/4}} \right) = 
O\left( \frac{n^{1+\eps}}{D^{(5/4+2\eps)k}s^{1/4}} \right) .
\] 
We multiply by the number of cells that the query visits, namely $O(D^{k+\eps})$, 
and add the cost $O(D^{k+\eps})$ of traversing these cells, for a total of
\[
O\left( D^{k+\eps} + \frac{n^{1+\eps}}{s^{1/3}} + \frac{n^{1+\eps}}{D^{(1/4+2\eps)k}s^{1/4}} \right) .
\]
In other words, we get the same bound as in (\ref{qtrade}), except for the
second term which is missing now (this term corresponds to querying at the
bottom-level nodes of the $r_0$-recursion on the wide triangles, which
is not needed when $s > n^2$, since these bottom-level subproblems
now have constant size).
Repeating the same analysis as above, we get the same bound for the query cost.

\paragraph{Handling the zero set.}
We next analyze the case where the query ray lies on the zero set.
In order to obtain the trade-off bounds for ray shooting within $Z(f)$,
we recall the multi-level data structure presented in Section~\ref{sec:onzf}. 
Each level in this data structure is either a one- or a two-dimensional search tree, 
where the dominating levels are those where we need to apply a planar decomposition 
over a set of planar regions (or in an arrangement of algebraic arcs) and preprocess it
into a structure that supports point-location queries. A standard property of multi-level 
range searching data structures is that the overall complexity of their storage (resp.,
query time) is governed by the level with dominating storage (resp., query time) bound,
up to a polylogarithmic factor~\cite{AE99}. Recall that in each level of our data structure 
we form a collection of canonical sets of the arcs in $\Gamma$, which are passed on to the 
next level for further processing. Our approach is to keep forming these canonical sets, 
where at the very last level we apply the ray-shooting data structure of Pellegrini~\cite{Pel}, 
as described above. Therefore the overall query cost (resp., storage and preprocessing complexity) 
is the sum of the query (resp., storage and preprocessing time) bounds over all canonical sets 
of arcs that the query reaches (resp., all the sets) at the last level.

We now sketch the analysis in more detail. In order to simplify the presentation, we consider 
one of the dominating levels, and describe the ray-shooting data structure at that level. 
As stated above, we build this data structure only at the very last level, but the analysis for
the dominating level subsumes the bounds for the last level, and thus for the entire multi-level 
data structure, up to a polylogarithmic factor. In such a scenario we have a set of algebraic 
arcs (or graphs of functions, or semi-algebraic regions represented by their bounding arcs),
which we need to preprocess for planar point location. This is done using the technique of 
$(1/r)$-cuttings (see~\cite{CF-90}), which forms a decomposition of the plane into $O(r^2)$ 
pseudo-trapezoidal cells, each meeting at most $n/r$ arcs (the ``conflict list'' of the cell). 
The overall storage complexity is thus $O(nr)$. More precisely, to achieve preprocessing time 
close to $O(nr)$, one needs to use so-called \emph{hierarchical-cuttings} (see~\cite{Mat} 
and also \cite{AES}), in which we construct
a hierarchy of cuttings using a constant value $r_0$ as the cutting parameter, instead of the
nonconstant $r$. Using this approach, both storage and preprocessing cost are 
$O(n r^{1+\eps})$ for any $\eps>0$. 
Let $s$ be our storage parameter as above, so we want to choose $r$ such that $s = r n$.
Thus we obtain that each cell of the cutting meets at most $n^2/s$ arcs.
Following our approach above, for each cell of the cutting, the amount of allocated storage is $s/r^2 = n^2/s$.
We are now ready to apply Pellegrini's data structure, leading to a query time of $O\left(\frac{n^{3/2+\eps}}{s^{3/4}}\right)$.
Integrating this bound into the query time in~(\ref{qtrade}), we recall that at each level $0 \le j \le k$ the actual
storage parameter is $O(s/D^{3j})$, and the number of triangles at hand is $O(n/D^{2j})$. 
We now need to sum the query bound over all $O(D^j)$ cells reached by the query at the $j$th level, 
and over all $j$. We thus obtain an overall bound of
$$
O\left(D^k \frac{(n/D^{2k})^{3/2+\eps}}{(s/D^{3k})^{3/4}} \right) =
O\left(\frac{n^{3/2+\eps} D^{k/4} }{s^{3/4}} \right) .
$$
This is exactly the second term in~(\ref{qtrade}). Therefore adding the query time 
for ray shooting on $Z(f)$ does not increase the asymptotic bound in~(\ref{qtrade}).

We comment that the overall storage and preprocessing time is $O(s^{1+\eps})$ (see our discussion below).
We also comment that the query bound we obtained applies when $n \le s \le n^2$. When $s$ exceeds $n^2$, 
every cell of the cutting has a conflict list of $O(1)$ elements, which the query can handle in 
brute-force. This immediately brings the query time, for queries on the zero set, to $O(n^{\eps})$.

\paragraph{Wrapping up.}
In summary, our analysis implies that the query bound $Q(n,s)$ satisfies:
\begin{equation}
  \label{eq:trade-q}
  Q(n,s) = \begin{cases}
    O\left( \frac{n^{5/4+\eps}}{s^{1/2}} \right) , & s = O(n^{3/2+\eps}) , \\
    O\left( \frac{n^{4/5+\eps}}{s^{1/5}} \right) , & s = \Omega(n^{3/2+\eps}) .
  \end{cases}
\end{equation}
We recall that the overall storage (and preprocessing) is $O(s^{1+\eps})$, 
since we allocate to each subproblem, at any level $j$, the storage parameter $s/D^{3j}$,
thus at each fixed level the total storage (and preprocessing) complexity is $O(s^{1+\eps})$,
and since there are only logarithmically many levels, the overall storage (and preprocessing)
is $O(s^{1+\eps})$ as well, for a slightly large $\eps$.

Note that for the threshold $s\approx n^{3/2}$, both bounds yield a query cost of $O(n^{1/2+\eps})$.
Note also that in the extreme cases $s = n^4$, $s = n$
(extreme for the older `four-dimensional' tradeoff), we get the 
respective older bounds $O(n^\eps)$ and $O(n^{3/4+\eps})$ for the query time.
In this case, when either $s = n$ and $s = n^4$ 
we have $D^{k} = O(1)$, implying that we handle all the narrow triangles at the root of the recursion tree,
that is, we use the technique of Pellegrini~\cite{Pel} once altogether.
Informally, the bound in~(\ref{eq:trade-q})
`pinches' the tradeoff curve and pushes it down. The closer $s$ is to 
$\Theta(n^{3/2+\eps})$, the more significant is the improvement.
See Figure~\ref{tradeoff}.

\paragraph{Processing $m$ queries.}
The improved tradeoff in (\ref{eq:trade-q})
implies that the overall cost of processing $m$ queries with $n$ input 
triangles, including preprocessing cost, is 
\[
O(s^{1+\eps} + mQ(n,s)) =
\begin{cases}
O\left( s^{1+\eps} + \frac{mn^{5/4}}{s^{1/2}} \right) , & s = O(n^{3/2+\eps}) , \\
O\left( s^{1+\eps} + \frac{mn^{4/5}}{s^{1/5}} \right) , & s = \Omega(n^{3/2+\eps}) .
\end{cases}
\]
To balance the terms in the first case we choose $s = m^{2/3}n^{5/6}$; 
this choice satisfies $s = O(n^{3/2+\eps})$ when $m\le n$.
To balance the terms in the second case we choose $s = m^{5/6}n^{2/3}$; 
this choice satisfies $s = \Omega(n^{3/2+\eps})$ when $m\ge n$.
Recall also that $s$ has to be in the range between $n$ and $n^4$. 
So in the first case we must have $m^{2/3}n^{5/6} \ge n$, or $m\ge n^{1/4}$.
Similarly, in the second case we must have $m^{5/6}n^{2/3} \le n^4$, or $m\le n^{4}$.
We adjust the bounds, allowing also values of $m$ outside this range, by adding the
near-linear terms, which dominate the bound for such off-range values of $m$. We thus get
\begin{corollary}
  \label{cor:queries}
  We can process $m$ ray-shooting queries on $n$ triangles so that the total cost is
  \begin{equation}
    \label{eq:trade-queries}
    \max\Bigl\{ O(m^{2/3+\eps}n^{5/6+\eps} + n^{1+\eps}),\; O(n^{2/3+\eps}m^{5/6+\eps} + m^{1+\eps}) \Bigr\} .
  \end{equation}
\end{corollary}

\section{Other applications} \label{sec:other}

\subsection{Detecting, counting or reporting line intersections in $\reals^3$}

It is more convenient, albeit not necessary, to consider the
bichromatic version of the problem, in which we are given a set $R$ of
$n$ red lines and a set $B$ of $n$ blue lines in $\reals^3$, and 
the detection problem asks whether there exists a pair of intersecting 
lines in $R\times B$. Similarly, the counting problem asks for the 
number of such intersecting pairs, and the reporting problem asks for
reporting all these pairs.

An algorithm that solves the detection problem in $O(n^{3/2+\eps})$ time 
is easily obtained by regarding the problem as a special degenerate
(and much simpler) instance of the ray shooting problem, in which
we regard the, say red lines as degenerate triangles (unbounded and 
of zero area), construct the data structure of Section~\ref{sec:shoot}
and query it with each of the blue lines. There exists a red-blue
pair of intersecting lines if and only if at least one query has
a positive outcome---the corresponding blue query line hits a red line.

Since there are no wide triangles in this special variant, there is
no need to construct the auxiliary data structure for wide triangles,
as in Section~\ref{sec:wide}, and we simply construct the recursive
hierarchy of polynomial partitions, where each cell in each subproblem 
is associated with the set of red lines that cross it. A blue query line
$\ell$ is propagated through the cells that it crosses until it reaches
bottom-level cells, and we check, in each such cell, whether $\ell$
intersects any of the $O(1)$ red lines associated with the cell.

Handling lines that lie fully in the zero set $Z(f)$ is also an easy 
task (which can be performed using planar segment-intersection range 
searching, which also supports counting queries); further details are omitted.

Both correctness and runtime analysis follow easily, as special
and simpler instances of the analysis in Section~\ref{sec:shoot}.
Note that here we do not face the issue of non-disjointness of canonical sets
of wide triangles, which has prevented us from extending the technique
to segment-triangle intersection counting problems; see Section~\ref{sec:seg}.

\subsection{Computing the intersection of two polyhedra} \label{2poly}

Let $K_1$ and $K_2$ be two polyhedra in 3-space, not necessarily convex,
each with $n$ edges (so the number of vertices and faces of each of them
is $O(n)$). The goal is to compute their intersection $K := K_1\cap K_2$
in an output-sensitive manner. We note that computing the union
$K_1\cup K_2$ can be done using a very similar approach, within 
the same time bound.

While there are additional steps in the algorithm that construct
a representation of $K$ as a three-dimensional body, we will restrict
here the presentation to the part that computes $\bd K$ from $\bd K_1$
and $\bd K_2$. Each face of $\bd K$ is a connected portion of a face 
of $\bd K_1$ or of $\bd K_2$, each edge is either a connected portion
of an edge of $\bd K_1$ or of $\bd K_2$, or a connected portion of
the intersection of a face of $\bd K_1$ and a face of $\bd K_2$.
Finally, each vertex of $\bd K$ is either a vertex of $\bd K_1$ 
or of $\bd K_2$, or an intersection of an edge of one of these 
polyhedra with a face of the other. Note that not every vertex of $K_1$
or of $K_2$ is necessarily a vertex of $K$, but every edge-face
intersection is a vertex of $K$.

The main step of the algorithm is to compute the vertices
of $\bd K$, from which the other features of $\bd K$ are fairly standard to construct,
see, e.g.,~\cite{Pel} where the graph of the edges of $K$ is constructed by
a tracing procedure~\cite{MS-85}, given the vertices of $\bd K$. 
We iterate over the edges of $\bd K_1$, and compute the intersections
of each such edge with the faces of $\bd K_2$, using the algorithm in Theorem~\ref{thm:seg_intersection}.
We apply a symmetric procedure to compute the intersections of each edge of $\bd K_2$
with the faces of $\bd K_1$. Collectively, these intersections are
the vertices of $K$ of the second type (edge-face intersection
vertices). The cost of this step is $O(n^{3/2+\eps} + k\log n)$, where
$k$ is the number of edge-face intersections: 
we preprocess the $O(n)$ faces of, say $K_1$, and query with the $O(n)$
edges of $K_2$, which overall takes 
$O(n\cdot n^{1/2+\eps} + k\log n) = O(n^{3/2+\eps} + k\log n)$ time.
Then, applying the tracing procedure in~\cite{MS-85} takes an additional 
cost of $O(k \log{k})$.

This gives us all the edge-face intersection vertices. The other vertices
of $K$ are vertices of $K_1$ or of $K_2$, and finding these vertices is
done as follows.
If such a vertex $v$, say of $K_1$,
is incident to an edge $e$ of $K_1$ that intersects some face of $K_2$, 
then it is easy to determine whether $v\in K_2$ (and thus in $K$).
Otherwise, we collect, by a simple graph traversal, a maximal cluster 
of vertices of $K_1$ that are connected by edges that have no intersection 
with $\bd K_2$. The vertices in such a cluster are either all inside $K_2$ 
(and in $K$) or all outside $K_2$ (and thus not in $K$). If the cluster
consists of all vertices of $K_1$ then either $K_1$, $K_2$ are disjoint, 
or one contains the other. In such a case, we only need to test, in $O(n)$ 
time, if there is a vertex from one polyhedron that is contained in the other.
Otherwise, we determine the status of the cluster (inside / outside $K$) by
examining the edges that connect vertices from the cluster to vertices not
in the cluster. By iteratively repeating this step, we construct all such
clusters, from which we obtain all the vertices of $K_1$ and of $K_2$
the lie in $K$.

In summary we obtain:

\begin{corollary}
  \label{cor:intersection_poly}
  Given two arbitrary polyhedra $K_1$ and $K_2$ each of complexity $O(n)$,
  the intersection $K_1 \cap K_2$ can be computed in time $O(n^{3/2+\eps} + k \log n)$,
  where $k$ is the size of the intersection.  
\end{corollary}
As discussed in the introduction, the overhead term in Pellegrini's 
algorithm~\cite{Pel} is $O(n^{8/5+\eps})$. 

\subsection{Output-sensitive construction of an arrangement of triangles}

Let $\T$ be a set of $n$ possibly intersecting triangles in $\reals^3$,
let $\A = \A(\T)$ denote their arrangement, and let $k$ denote its
complexity, which, as in Section~\ref{2poly}, we measure by the 
number of its vertices, as the number of its other features (edges, 
faces, and cells) is proportional to $k$. The goal
is to construct $\A$ in an output-sensitive manner with a small,
subquadratic overhead. Pellegrini~\cite{Pel} gave such an algorithm
that runs in $O(n^{8/5+\eps} + k\log k)$, and the algorithm that we
present here reduces the overhead to $O(n^{3/2+\eps})$ time.

As in the previous subsection, we focus on the main step of the 
algorithm that constructs the features of $\A$ (vertices, edges, 
and faces) on each triangle of $\T$. We will only briefly discuss 
the complementary part, which constructs the three-dimensional cells 
of $\A$ and establishes the connections between the various features 
on the boundary of each cell. Albeit not trivial, this latter step
uses standard techniques, follows the approach in \cite{Pel} and 
in other works, and does not increase the overhead cost of the algorithm.
 
Fix a triangle $\Delta\in\T$. We first construct the set of intersection
segments $\Delta\cap\Delta'$, for $\Delta'\in\T\setminus\{\Delta\}$.
We observe that, for any such segment $e = \Delta\cap\Delta'$, each 
endpoint of $e$ is either a vertex of $\Delta$, 
or an intersection of an edge of one triangle with the other triangle.

We therefore take the collection of the $3n$ edges of the triangles
of $\T$, and, for each such edge $e$, apply Theorem~\ref{thm:seg_intersection},
which reports all $k_e$ triangles that $e$ meets. This identifies all the
intersection segments $\Delta\cap\Delta'$. We then take all the
intersection segments within a fixed triangle $\Delta$, and run
a sweepline procedure within $\Delta$ to obtain the portion of $\A$
on $\Delta$. Gluing these portions to each other, and
some additional steps, complete the construction of $\A$.

\section{Conclusion}

In this paper we have managed to improve the performance of ray shooting
amid triangles in three dimensions, as well as of several related problems.
The improvement is based on the polynomial partitioning technique of Guth.
The improvement is most significant when the storage is about $n^{3/2}$ and 
the query takes about $n^{1/2}$ time, but one gets an improvement for all values
of the storage between $n$ and $n^4$, except at the very ends of this range. 
This is a significant improvement, the first in nearly 30 years, in this basic problem.

There are several open questions that our work raises. First, the improvement
for the special values of $O(n^{3/2+\eps})$ storage and $O(n^{1/2+\eps})$ 
query time seems too specialized, and one would like to get similar
improvements for all possible values of the storage, ideally obtaining
query time of $O(n^{1+\eps}/s^{1/3})$, where $s$ is the storage allocated 
to the structure, as in the case of ray shooting amid planes. Alternatively,
can one establish a lower-bound argument that shows the limitations of our technique?

Another open issue follows from our current inability to extend the technique to 
counting queries, due to the fact that the canonical sets that we collect 
during a query are not necessarily pairwise disjoint. It would be interesting
to obtain such an extension, or, alternatively, to establish a gap between
the performances of the counting and reporting versions of the segment 
intersection query problem. 

Finally, could one obtain similar bounds for non-flat input objects? for shooting
along non-straight curves? It would also be interesting to find 
additional applications of the general technique developed in this paper.

\paragraph{Acknowledgements.}
We wish to thank Pankaj Agarwal for the useful interaction concerning 
certain aspects of the range searching problems encountered in this work.

\end{document}